\documentclass[preprint2]{proto}
\usepackage{times}
\newcommand{\refs}{\par\noindent\hangindent=1pc\hangafter=1}
\begin{document}

\title{\textbf{\LARGE The Formation of Brown Dwarfs}}

\author {\textbf{\large Anthony Whitworth}}
\affil{\small\em Cardiff University}
\author {\textbf{\large Matthew R. Bate}}
\affil{\small\em University of Exeter}
\author {\textbf{\large {\AA}ke Nordlund}}
\affil{\small\em University of Copenhagen}
\author {\textbf{\large Bo Reipurth}}
\affil{\small\em University of Hawai'i}
\author {\textbf{\large Hans Zinnecker}}
\affil{\small\em Astrophysikalisches Institut, Potsdam}

\begin{abstract}
\begin{list}{ } {\rightmargin 1in}
\baselineskip = 11pt
\parindent=1pc
{\small We review five mechanisms for forming brown dwarfs: (i) turbulent fragmentation of molecular clouds, producing very low-mass prestellar cores by shock compression; (ii) collapse and fragmentation of more massive prestellar cores; (iii) disc fragmentation; (iv) premature ejection of protostellar embryos from their natal cores; and (v) photo-erosion of pre-existing cores overrun by HII regions. These mechanisms are not mutually exclusive. Their relative importance probably depends on environment, and should be judged by their ability to reproduce the brown-dwarf IMF, the distribution and kinematics of newly formed brown dwarfs, the binary statistics of brown dwarfs, the ability of brown dwarfs to retain discs, and hence their ability to sustain accretion and outflows. This will require more sophisticated numerical modelling than is presently possible, in particular more realistic initial conditions and more realistic treatments of radiation transport, angular momentum transport and magnetic fields. We discuss the minimum mass for brown dwarfs, and how brown dwarfs should be distinguished from planets. 
 \\~\\~\\~}
 
\end{list}
\end{abstract}

\section{\textbf{INTRODUCTION}} \label{SEC:INTR}%

The existence of brown dwarfs was first proposed on theoretical grounds by {\it Kumar} (1963) and {\it Hayashi and Nakano} (1963). However, more than three decades then passed before brown dwarfs were observed unambiguously ({\it Rebolo et al.}, 1995; {\it Nakajima et al.}, 1995; {\it Oppenheimer et al.}, 1995). Brown dwarfs are now observed routinely, and it is therefore appropriate to ask how brown dwarfs form, and in particular to ascertain (a) whether brown dwarfs form in the same way as H-burning stars, and (b) whether there is a clear distinction between the mechanisms that produce brown dwarfs and those that produce planets.

In \S \ref{SEC:STAR} we argue that the mechanisms forming brown dwarfs are no different from those forming low-mass H-burning stars, on the grounds that the statistical properties of brown dwarfs (mass function, clustering properties, kinematics, binary statistics, accretion rates, etc.) appear to form a smooth continuum with those of low-mass H-burning stars. Understanding how brown dwarfs form is therefore the key to understanding what determines the minimum mass for star formation. In \S \ref{SEC:PHYS} we review the basic physics of star formation, as it applies to brown dwarfs, and derive key analytic results, in particular the minimum mass for opacity-limited fragmentation in various different formation scenarios. We have deliberately assembled most of the mathematical analysis in this one section. In \S \ref{SEC:TURB} to \S \ref{SEC:EROS} we consider five different mechanisms that may be involved in the formation of brown dwarfs. \S \ref{SEC:TURB} explores the possibility that turbulent fragmentation of molecular clouds produces prestellar cores of such low mass that inevitably they collapse to form brown dwarfs. \S \ref{SEC:FRAG} considers more massive prestellar cores, and the possibility that they fragment dynamically as they collapse, thereby spawning protostars with a range of masses. The collapse of such cores ceases when the gas starts to heat up (due to adiabatic compression) and/or when rotation becomes important. Unfortunately, the interplay of dynamics and adiabatic compression, whilst likely to play a critical role (e.g. {\it Boss et al.}, 2000), is hard to analyse. On the other hand rotational effects can be analysed systematically, and therefore \S \ref{SEC:DISC} explores the formation of brown dwarfs by gravitational instabilities in discs, considering first isolated relaxed discs, then unrelaxed discs, and finally interacting discs in clusters. \S \ref{SEC:EJEC} reviews the process of competitive accretion, which determines how an ensemble of protostellar embryos evolves to populate higher masses; and then considers the $N$-body processes which (a) may eject brown dwarfs and low-mass stars from their natal cores, thereby terminating accretion and effectively capping their masses, and (b) may influence the binary statistics and clustering properties of brown dwarfs. \S \ref{SEC:EROS} considers the formation of brown dwarfs by photo-erosion of pre-existing cores which are overrun by HII regions. \S \ref{SEC:SIMS} presents numerical simulations of the birth of a whole star cluster in a large protocluster core; in these simulations, the mechanisms of \S \ref{SEC:FRAG} (collapse and fragmentation), \S \ref{SEC:DISC} (disc fragmentation) and \S \ref{SEC:EJEC} (premature ejection) all occur simultaneously and in tandem, and the collective properties of brown dwarfs formed can be extracted for comparison with observation. In \S \ref{SEC:CONC} we summarise the principal conclusions of this review.

Comprehensive discussions of the observational properties of brown dwarfs, and how they may constrain formation mechanisms, are contained in the chapters by {\it Luhman et al.} (which deals with the entire observational picture) and {\it Burgasser et al.} (which deals specifically with the issue of multiplicity). Therefore we offer only a brief summary of the observations in \S \ref{SEC:STAR}. Likewise, a critique of simulations of core fragmentation is given in the chapter by {\it Goodwin et al.}; simulations of disc fragmentation are discussed and compared in detail in the chapter by {\it Durisen et al.};  and the origin of the IMF is treated in the chapter by {\it Bonnell et al.}, so we have limited our consideration to those aspects which pertain specifically to the origin of brown dwarfs.

\section{\textbf{EVIDENCE THAT BROWN DWARFS FORM LIKE LOW-MASS H-BURNING STARS}}%
                                                                \label{SEC:STAR}%

We shall assume that brown dwarfs form in the same way as H-burning stars, i.e. on a dynamical timescale, by gravitational instability, and with initially uniform elemental composition (reflecting the composition of the interstellar matter out of which they form). (By implication, we distinguish {\it brown dwarfs} from {\it planets}, a term which we will here reserve for objects which form on a much longer timescale, by the amalgamation of a rocky core and -- if circumstances allow -- the subsequent accretion of a gaseous envelope. This results in planets having an initially fractionated elemental composition with an overall deficit of light elements.) If this is the correct way to view the formation of brown dwarfs, then brown dwarfs should not be distinguished from stars. Many stars fail to burn helium, and most fail to burn carbon, without forfeiting the right to be called stars. The reason for categorising brown dwarfs as stars is that the statistical properties of brown dwarfs appear to form a continuum with those of low-mass H-burning stars (and not with those of high-mass planets).
 
\subsection{\textbf{The IMF, clustering statistics, velocity dispersion}}

The initial mass function (IMF) is apparently continuous across the hydrogen-burning limit at $\sim 0.075\,{\rm M}_{_\odot}$, in the Trapezium cluster ({\it Slesnick et al.}, 2004), in the $\sigma$ Orionis cluster ({\it Bejar et al.}, 2001), in Taurus ({\it Luhman}, 2004), in IC348 ({\it Luhman et al.}, 2003), in the Pleiades ({\it Moraux et al.}, 2003), in the field stars of the disc ({\it Chabrier}, 2003) and even possibly in the halo ({\it Burgasser et al.}, 2003b). If the IMF is fitted by a power law across the H-burning limit, $d{\cal N}/dM \propto M^{-\alpha}$, estimates of $\alpha$ fall in the range 0.4 to 0.8 (e.g. {\it Moraux et al.}, 2003). The IMF appears to extend down to a few Jupiter masses (e.g. {\it Zapatero Osorio et al.}, 2002; {\it McCaughrean et al.}, 2002; {\it Lucas et al.}, 2005). The continuity of the IMF across the H-burning limit is not surprising, since the processes which determine the mass of a low-mass star are presumed to occur at relatively low densities ($\rho \la 10^{-8}\,{\rm g}\,{\rm cm}^{-3}$) and temperatures ($T \la 2000\,{\rm K}$), long before the material involved knows whether it will reach sufficiently high density ($\rho \ga 1\,{\rm g}\,{\rm cm}^{-3}$) to be supported in perpetuity by electron degeneracy pressure before or after it reaches sufficiently high temperature ($T \ga 10^7\,{\rm K}$) to burn hydrogen.

In the Trapezium cluster, and in Taurus, brown dwarfs appear to be homogeneously mixed with H-burning stars ({\it Lucas and Roche}, 2000; {\it Brice\~no et al.}, 2002; {\it Luhman}, 2004), and in Chamaeleon I their kinematics are also essentially indistinguishable ({\it Joergens}, 2006b). Although they have been searched for -- as possible signatures of formation by ejection -- neither a greater velocity dispersion of brown dwarfs in very young clusters, nor a diaspora of brown dwarfs around older clusters, has been found.

\subsection{\textbf{Binary statistics}}

Multiple systems involving brown dwarf secondaries can be categorised based on whether the primary is a Sun-like star or another brown dwarf.

{\em Sun-like primaries.} Sun-like stars seldom have brown-dwarf companions. At close separations ($\la 5\,{\rm AU}$), the frequency of companions with masses in the range $0.01\;{\rm to}\;0.1\,{\rm M}_{_\odot}$ is $\,\sim 0.5\%$ ({\it Marcy and Butler}, 2000); this figure is known rather accurately due to the numerous Doppler surveys aimed at detecting extrasolar planets.  Since the frequency of companions outside this mass range (both exoplanets below, and H-burning stars above) is much higher, the paucity of brown-dwarf companions is termed the `brown dwarf desert'. However, the chapter by {\it Luhman et al.} points out that the paucity of close brown-dwarf companions to Sun-like stars may simply reflect the overall paucity of brown dwarfs. At larger separations ($\ga 100\,{\rm AU}$), only about 20 brown-dwarf companions to Sun-like stars have been found to date, indicating that these systems are also rare (frequency $\sim 1\;{\rm to}\;2\,\%$), though this estimate is based on more limited statistics (e.g. {\it Gizis et al.}, 2001; {\it McCarthy and Zuckerman}, 2004).

{\em Brown-dwarf primaries.} BD-BD binary systems (and binary systems with very low-mass H-burning primaries) have an observed multiplicity of $\sim 10-20\%$ for separations greater than $\sim 2\,{\rm AU}$ ({\it Bouy et al.}, 2003; {\it Burgasser et al.}, 2003a; {\it Close et al.}, 2003; {\it Gizis et al.}, 2003), and all but five of the $\sim 75$ known BD-BD binaries have separations less than $\sim 20\,{\rm AU}$ (see Table 1 in the chapter by {\it Burgasser et al.}).  Below $\sim 2\,{\rm AU}$, the multiplicity is unknown because most surveys to date have been imaging surveys that cannot resolve close systems.  Only a few spectroscopic BD-BD binaries have been discovered ({\it Basri and Martin}, 1999; {\it Stassun et al.}, 2006; {\em Kenyon et al.}, 2005).  Some authors speculate that the overall multiplicity might be $\sim 30-50\%$, based on the positions of brown dwarfs in colour-magnitude diagrams ({\it Pinfield et al.}, 2003), and the statistics of radial velocity variations ({\it Maxted and Jeffries}, 2005). However, {\it Joergens} (2006a) has examined the radial velocities of ten BDs and very low-mass H-burning stars, and finds no binaries with separations less than $\sim 0.1\,{\rm AU}$, and only two objects with variability on timescales greater than 100 days that might indicate companions at greater than $\sim 0.2\,{\rm AU}$.  Thus, the peak in the separation distribution for BD-BD binaries is likely to be at $\sim 1\;{\rm to}\;4\,{\rm AU}$. In contrast, the distribution of separations in binaries with Sun-like primaries peaks at $\sim 30\,{\rm AU}$ ({\it Duquennoy \& Mayor}, 1991). The mass ratio distribution also seems to be dependent on primary mass, with BD-BD binaries having a distribution which peaks towards equal masses ($q_{_{\rm PEAK}}\sim 1$), whilst binaries with Sun-like primaries have $q_{_{\rm PEAK}} \sim 0.3$. The implication is that, as the primary mass decreases, (i) the multiplicity decreases, (ii) the distribution of semi-major axes shifts to smaller separations and becomes narrower (logarithmically), and (iii) the distribution of mass ratios shifts towards unity -- with these trends all continuing across the divide between H-burning stars and brown dwarfs.

{\em Exotica.} Finally, we note that some brown dwarfs are components in more complex systems.  There are at least six systems currently known in which an H-burning star is orbited at large distance ($\ga 50\,{\rm AU}$) by a BD-BD binary. Indeed, preliminary results suggest that wide brown-dwarf companions to H-burning stars are 2 to 3 times more likely to be in a close BD-BD binary than are field brown dwarfs ({\it Burgasser et al.}, 2005). Cases also exist in which a close binary with H-burning components is orbited by a wide brown dwarf, while {\it Bouy et al.} (2005) have recently reported the discovery of what is likely to be a triple brown dwarf system. However, the statistics of these exotic systems are currently too small to interpret quantitatively.

\subsection{\textbf{Discs, accretion and outflows}}

Young brown dwarfs are observed to have infrared excesses indicative of circumstellar discs, like young H-burning stars ({\it Muench et al.}, 2001; {\it Natta \& Testi}, 2001; {\it Jayawardhana et al.}, 2003; {\it Mohanty et al.}, 2004). Brown-dwarf disc lifetimes are estimated to be 3 to $10\,{\rm Myr}$, again like H-burning stars. From their H$\alpha$ emission-line profiles, there is evidence for ongoing magnetospheric accretion onto brown dwarfs ({\it Scholz and Eisl\"offel}, 2004), and the inferred accretion rates form a continuous distribution with those for H-burning stars, fitted approximately by $\dot{M} \sim 10^{-8}\,{\rm M}_{_\odot}\,{\rm yr}^{-1}\,\left(M/{\rm M}_{_\odot}\right)^2$ ({\it Muzerolle et al.}, 2003, 2005). To sustain these estimated accretion rates, brown dwarfs only require rather low-mass discs $\sim 10^{-4}\,{\rm M}_{_\odot}$. Finally, the spectra of brown dwarfs also show forbidden emission lines suggestive of outflows like those from H-burning stars ({\it Fern\'andez and Comer\'on}, 2001; {\it Natta et al.}, 2004), and recently an outflow from a brown dwarf has been resolved spatially ({\it Whelan et al.}, 2005). Thus, in the details of their circumstellar discs, accretion rates and outflows, young brown dwarfs appear to mimic H-burning stars very closely, and to differ significantly only in scale.

\subsection{\textbf{Rotation and X-rays}}

The rotational properties of brown dwarfs also appear to connect smoothly with those of very low-mass H-burning stars (e.g. {\it Joergens et al.}, 2003). There is a decrease in the amplitude of periodic photometric variations with decreasing mass, presumably because the decreasing surface temperature leads to weaker coupling between the gas and the magnetic field and hence smaller spots. As with H-burning stars, brown dwarfs show evidence for braking by their accretion discs; those with strong accretion signatures (broad H$\alpha$ emission) are exclusively slow rotators.

X-ray emission is detected from M-type brown dwarfs, with properties very similar to the X-ray emission from very low-mass H-burning stars ({\it Preibisch et al.}, 2005), but none is detected from L-type brown dwarfs, which are too cool to support surface magnetic activity.

\subsection{\textbf{Synopsis}}

Given the continuity of statistical properties between brown dwarfs and H-burning stars, it is probably unhelpful to distinguish the formation of brown dwarfs from the formation of stars, and in the rest of this review we will only use the H-burning limit at $\sim 0.075 {\rm M}_{_\odot}$ as a reference point in the range of stellar masses. The D-burning limit at $\sim 0.012\,{\rm M}_{_\odot}$ falls in the same category. We will then define a star as any object forming on a dynamical timescale, by gravitational instability. With this definition, there may well be a small overlap between the mass range of stars and that of planets. Given that in the immediate future we are unlikely to know too much more than the masses and radii of the lowest-mass objects, and certainly not their internal composition, we will simply have to accept that there is a grey area in the range $0.001\,{\rm to}\,0.01\,{\rm M}_{_\odot}$ which may harbour both stars and planets, and possibly even hybrids.

It follows that understanding how brown dwarfs form is important, not just for its own sake, but because it is a key element in understanding why most stars have masses in the range $0.01\,{\rm M}_{_\odot}$ to $100\,{\rm M}_{_\odot}$ -- and hence why there are lots of hospitable stars like the Sun with long-lived habitable zones, and enough heavy elements to produce rocky planets and life. The high-mass cut-off is probably due to the fact that radiation pressure makes it hard to form the highest-mass stars. The low-mass cut-off is probably due to the opacity limit. By studying brown dwarf formation we seek to confirm and quantify the low-mass cut-off.

\section{\textbf{STAR FORMATION THERMODYNAMICS}} \label{SEC:PHYS} %

In this section we review the basic thermodynamics of gravitational collapse and fragmentation. The first three subsections deal -- respectively -- with 3-D collapse and hierarchical fragmentation (\S 3.1); 2-D one-shot fragmentation of a shock compressed layer (\S 3.2); and fragmentation of a disc (\S 3.3). We describe and contrast these different environments, and in each case estimate the minimum mass for star formation. In \S 3.3 we conclude that brown dwarf formation by fragmentation of discs around Sun-like stars is more likely in the cooler outer parts ($R \ga 100\,{\rm AU}$), and this may explain why brown dwarf companions to Sun-like stars almost always have large separations. In \S 3.4 we explain why impulsive compression does not promote cooling of an optically thick fragment. In \S 3.5 we suggest that close BD-BD binaries may be formed by secondary fragmentation promoted by the softening of the equation of state when H$_{_2}$ dissociates, and enhanced cooling due to the opacity gap. In \S 3.6 we speculate that close BD-BD binaries may form in the outer parts of discs around Sun-like stars.

We caution that analytic estimates cannot capture all the non-linear effects which are likely to occur, and are probably important, in a process as chaotic as star formation; they are therefore only indicative. A full understanding of any mode of star formation requires detailed simulations with all potentially influential physical effects included. However, as long as converged, robust simulations with all this physics properly included remain beyond the compass of current supercomputers, analytic estimates provide useful insights into the trends to be expected.

In \S 3.1 to \S 3.3, we will be mainly concerned with contemporary star formation in the disc of the Milky Way, and therefore with molecular hydrogen at temperatures $T \la 100\,{\rm K}$ where the rotational levels are not strongly excited. In this regime the adiabatic exponent is $\gamma \simeq 5/3$ and the isothermal sound speed is $a \simeq 0.06\,{\rm km}\,{\rm s}^{-1}\,(T/{\rm K})^{1/2}$. With these assumptions, our estimates will not apply to the hot gas ($T > 10^3\,{\rm K}$) where the equation of state is softened by effects like H$_{_2}$ dissociation (e.g. the inner regions of discs). We will also assume that the metallicity is approximately solar, and that the Rosseland- and Planck-mean opacities due to dust are (to order of magnitude) the same, $\bar{\kappa}_{_{\rm R}}(T) \simeq \bar{\kappa}_{_{\rm P}}(T) \simeq \kappa_{_1}(T/{\rm K})^\beta$, with $\kappa_{_1} = 10^{-3}\,{\rm cm}^2\,{\rm g}^{-1}$ and emissivity index $\beta = 2$ in the far-infrared and submillimetre. With this assumption, our estimates will again not apply to the hot gas ($T > 10^3\,{\rm K}$) where dust sublimates and the opacity falls abruptly with increasing temperature (before picking up at even higher temperatures, due to the H$^{^-}$ ion). In \S 3.4 to \S 3.6, we relax these assumptions.

\subsection{\textbf{3-D collapse and hierarchical fragmentation}}

In a uniform 3-D medium, an approximately spherical fragment of mass $M_{_{\rm F3}}$ will only condense out if it is sufficiently massive,
\begin{equation} \label{EQN:MJEANS3} 
M_{_{\rm F3}} \;>\; M_{_{\rm J3}}  \;\simeq\; \left[ \frac{\pi^5a^6}{36G^3\rho} \right]^{1/2} \,;
\end{equation}
or equivalently, if it is sufficiently small and dense,
\begin{eqnarray} \label{EQN:RHOJEANS3} 
\rho_{_{\rm F3}} & > & \rho_{_{\rm J3}}  \;\simeq\; \frac{\pi^5a^6}{36G^3M_{_{\rm F3}}^2} \,. 
\end{eqnarray}
(Subscript {\sc f} is for fragment, {\sc j} is for Jeans, and {\small 3} is for 3-D.) The timescale on which the fragment condenses out is 
\begin{equation} \label{EQN:TIME3} 
t_{_{\rm F3}} \simeq \left[\frac{3\pi}{32G\rho}\right]^{1/2} \left[1- \left(\frac{M_{_{\rm J3}}}{M_{_{\rm F3}}}\right)^{2/3}\right]^{-1/2} 
\end{equation}

The molecular-cloud gas from which stars are forming today in the Milky Way is expected to be approximately isothermal, with $T \sim 10\,{\rm K}$, as long as it can radiate efficiently via molecular lines and dust continuum. Therefore it has been argued, following {\it Hoyle} (1953), that star formation proceeds in molecular clouds by a process of hierarchical fragmentation in which an initially massive low-density cloud (destined to form a proto-cluster of stars) satisfies condition (\ref{EQN:MJEANS3}) and starts to contract. Once its density has increased by a factor $f^2$, $M_{_{\rm J3}}$ is reduced by a factor $f^{-1}$, and hence parts of the cloud can condense out independently, thereby breaking the cloud up into $\la f$ sub-clouds. Moreover, as long as the gas remains approximately isothermal, the process can repeat itself recursively, breaking the cloud up into ever smaller `sub-sub...sub-clouds'.

The process ends when the smallest sub-sub...sub-clouds are so optically thick, and/or collapsing so fast, that the $PdV$ work being done on them cannot be radiated away fast enough and they heat up. This process is presumed to determine the minimum mass for star formation (e.g. {\em Rees}, 1976; {\em Low and Lynden-Bell}, 1976), and is usually referred to as The Opacity Limit, but see {\em Masunaga and Inutsuka} (1999) for a more accurate discussion. To estimate $M_{_{\rm MIN3}}$ we first formulate the $PdV$ heating rate for a spherical fragment, neglecting the background radiation field,
\begin{equation} \label{EQN:HEAT3} 
{\cal H} \equiv -P\frac{dV}{dt} = -\frac{3M_{_{\rm F3}}a^2}{R_{_{\rm F3}}}\frac{dR_{_{\rm F3}}}{dt} \sim \frac{3M_{_{\rm F3}}a^2}{R_{_{\rm F3}}}\!\left[ \frac{GM_{_{\rm F3}}}{R_{_{\rm F3}}}\right]^{1/2}\!, 
\end{equation}
where in putting $dR_{_{\rm F3}}/dt \sim - (GM_{_{\rm F3}}/R_{_{\rm F3}})^{1/2}$ we are assuming the collapse is dynamical. By comparison, the maximum radiative luminosity of a spherical fragment is 
\begin{equation} \label{EQN:COOL3} 
{\cal L} \simeq \frac{4\pi R_{_{\rm F3}}^2\sigma_{_{\rm SB}}T^4}{\left(\bar{\tau}_{_{\rm R}}(T) + \bar{\tau}_{_{\rm P}}^{\;-1}(T)\right)} \,, 
\end{equation}
where the optical depths are given by $\bar{\tau}_{_{\rm R}}(T) \simeq \bar{\tau}_{_{\rm P}}(T) \simeq 3M_{_{\rm F3}}\kappa_{_1}(T/{\rm K})^2/4\pi R_{_{\rm F3}}^2$. 

If we follow {\em Rees} (1976) and assume that the fragment is marginally optically thick, we can put $\left(\bar{\tau}_{_{\rm R}}(T) + \bar{\tau}_{_{\rm P}}^{\;-1}(T)\right) \simeq 2$, and the requirement that ${\cal L} \ga {\cal H}$ then reduces to
\begin{eqnarray} \label{EQN:RHOCOOL3} 
\rho_{_{\rm F3}} \la \rho_{_{\rm C3}} & \simeq & \left[\frac{\pi^{29}}{2^23^55^6}\,\frac{\bar{m}^{24}a^{36}}{G^3M_{_{\rm F3}}^2c^{12}h^{18}}\right]^{1/7}, 
\end{eqnarray}
Conditions (\ref{EQN:RHOJEANS3}) and (\ref{EQN:RHOCOOL3}) require $\rho_{_{\rm J3}} < \rho_{_{\rm C3}}$ and hence
\begin{equation}
M_{_{\rm F3}} \ga M_{_{\rm MIN3}} \simeq \left[\frac{5^2\pi^2}{2^43^3}\right]^{1/4}\,\frac{m_{_{\rm PL}}^3}{\bar{m}^2}\,\left[\frac{a}{c}\right]^{1/2}.
\end{equation}
Here $m_{_{\rm PL}} = (hc/G)^{1/2} = 5.5 \times 10^{-5}\,{\rm g}$ is the Planck mass, and so $M_{_{\rm MIN3}}$ is essentially the Chandrasekhar mass times a factor $(a/c)^{1/2} \sim 10^{-3}$. We also note the relatively weak dependence of $M_{_{\rm MIN3}}$ on $T$ ($\propto T^{1/4}$) and the relatively strong dependence on $\bar{m}$ ($\propto \bar{m}^{-9/4}$). For contemporary local star formation we substitute $\bar{m} \simeq 4.0 \times 10^{-24}\,{\rm g}$ and $a \simeq 1.8 \times 10^4\,{\rm cm}\,{\rm s}^{-1}$ to obtain $M_{_{\rm MIN3}} \sim 0.004\,{\rm M}_{_\odot}$. 

In general, the limiting fragment will not necessarily be marginally optically thick, but it is trivial to substitute for $\bar{\tau}_{_{\rm R}}(T)$ and $\bar{\tau}_{_{\rm P}}(T)$, and it turns out that, {\em for contemporary star formation}, the value of $M_{_{\rm MIN3}}$ is unchanged. This is because -- coincidentally -- the limiting fragment in contemporary star formation {\em is} marginally optically thick.

There are, however, some serious problems with 3-D hierarchical fragmentation. There is no conclusive evidence that it operates in nature, and nor does it seem to occur in numerical simulations of star formation. This is because a proto-fragment inevitably condenses out more slowly than the larger structure of which it is part, by virtue of the fact that it is, at every stage, less Jeans unstable than that larger structure (see Eqn. \ref{EQN:TIME3}). Therefore the proto-fragment is very likely to be merged with other nearby fragments before its condensation becomes nonlinear. In addition, even if a proto-fragment starts off with mass $\sim M_{_{\rm J3}}$, it will subsequently increase its mass by a large factor through accretion, before its condensation becomes nonlinear. Finally, individual fragments will be back-warmed by the ambient radiation field from other cooling fragments, which in principle fill a significant fraction of the celestial sphere, and again this will tend to increase $M_{_{\rm MIN3}}$.

\subsection{\textbf{2-D one-shot fragmentation of a shocked layer}}

In fact, 3-D hierarchical fragmentation may be an inappropriate paradigm for star formation in molecular clouds. There is growing evidence that, once a molecular cloud is assembled, star formation proceeds very rapidly, essentially `in a crossing time' ({\it Elmegreen}, 2000). In this scenario, star formation occurs in molecular clouds only where two or more turbulent flows of sufficient density collide with sufficient ram pressure to produce a shock-compressed layer or filament which then fragments to produce prestellar cores; in cases where flows converge simultaneously from several directions, isolated cores may even form (see {\it Padoan and Nordlund}, 2002, 2004; and  \S \ref{SEC:TURB}). 

A basic model of this scenario can be constructed with relatively few parameters by considering two flows having pre-shock density $\rho$, colliding at relative speed $v$, to produce a shock-compressed layer. If the effective isothermal sound speed in the resulting layer is $a$, the density is $\sim \rho (v/a)^2$. The layer is initially contained by the ram pressure of the inflowing gas, and until it fragments it has a rather flat density profile. It fragments at time $t_{_{\rm F2}}$, whilst it is still accumulating,  and the fastest growing fragments have mass $M_{_{\rm F2}}$, radius $R_{_{\rm F2}}$ (in the plane of the layer) and half-thickness $Z_{_{\rm F2}}$ (perpendicular to the plane of the layer) given by
\begin{eqnarray}
t_{_{\rm F2}} & \sim & (2 a / \pi G \rho v)^{1/2} \,, \\ \label{EQN:MFRAG2}
M_{_{\rm F2}} & \sim & (8 a^7 / \pi G^3 \rho v)^{1/2} \,, \\
R_{_{\rm F2}} & \sim & (2 a^3 / \pi G \rho v)^{1/2} \,, \\
Z_{_{\rm F2}} & \sim & (8 a^5 / \pi G \rho v^3)^{1/2}
\end{eqnarray}
(see {\it Whitworth et al.}, 1994a, 1994b). We note (a) that this mode of fragmentation is `2-D' because the motions which assemble a fragment out of the shock-compressed layer are initially largely in the plane of the layer; (b) that it is `one-shot' in the sense of not being hierarchical; (c) that the fragments are initially flattened objects ($R_{_{\rm F2}} / Z_{_{\rm F2}} \sim v / 2a \gg 1$); (d) that $M_{_{\rm F2}}$ is not simply the standard 3-D Jeans mass ($M_{_{\rm J3}}$, Eqn. \ref{EQN:MJEANS3}) evaluated at the post-shock density and velocity dispersion -- it is larger by a factor $\sim (v/a)^{1/2}$; and (e) that in reality the colliding flows will contain density substructure which acts as seeds for fragmentation and gives rise to a range of $M_{_{\rm F2}}$ values.

From Eqn. (\ref{EQN:MFRAG2}), we see that the fragment mass decreases monotonically with the mass-flux in the colliding flows, $\rho v$. As in  hierarchical fragmentation, a fragment in a shock-compressed layer will only condense out if it is able to remain approximately isothermal by radiating efficiently. The $PdV$ heating rate for a flattened fragment in a layer is
\begin{equation} \label{EQN:HEAT2}
{\cal H} \equiv -P\frac{dV_{_{\rm F2}}}{dt} \simeq \frac{\rho v^2}{4}\frac{2\pi R_{_{\rm F2}}^2 Z_{_{\rm F2}}}{t_{_{\rm F2}}} \simeq \frac{2a^5}{G}.
\end{equation}
The radiative cooling rate of the fragment is
\begin{equation}
{\cal L} \simeq \frac{2 \pi R_{_{\rm F2}}^2 \sigma_{_{\rm SB}} T^4}{\left(\bar{\tau}_{_{\rm R}}(T)+\bar{\tau}_{_{\rm P}}^{\;-1}(T)\right)} \simeq \frac{8\pi^5\bar{m}^4a^{11}/15c^2h^3G\rho v}{\left(\bar{\tau}_{_{\rm R}}(T)+\bar{\tau}_{_{\rm P}}^{\;-1}(T)\right)},
\end{equation}
where the optical depths are now given by $\bar{\tau}_{_{\rm R}}(T) = \bar{\tau}_{_{\rm P}}(T) = \left(2a\rho v/\pi G\right)^{1/2}\kappa_{_1}(T/{\rm K})^2$. The requirement that ${\cal L} \ga {\cal H}$ then reduces to a limit on the mass flux in the colliding flows,
\begin{equation}
\rho v \la \frac{4\pi^5\bar{m}^4a^6/15c^2h^3}{\left(\bar{\tau}_{_{\rm R}}(T)+\bar{\tau}_{_{\rm P}}^{\;-1}(T)\right)},
\end{equation}
If we assume the fragment is marginally optically thick, and set $\left(\bar{\tau}_{_{\rm R}}(T)+\bar{\tau}_{_{\rm P}}^{\;-1}(T)\right) \simeq 2$, we obtain
\begin{equation} \label{EQN:2DMARG}
M_{_{\rm F2}} \ga M_{_{\rm MIN2}} \simeq \frac{(30)^{1/2}}{\pi^3} \frac{m_{_{\rm PL}}^3}{\bar{m}^2} \left[\frac{a}{c}\right]^{1/2};
\end{equation}
and for contemporary local star formation, this gives $M_{_{\rm MIN2}} \sim 0.001\,{\rm M}_{_\odot}$. Once again, if we treat the completely general case by including the optical-depth terms, we obtain essentially the same value for $M_{_{\rm MIN2}}$, because -- purely coincidentally -- the limiting mass for contemporary local star formation is again marginally optically thick.

Although one-shot 2-D fragmentation of a shock-compressed layer and 3-D hierarchical fragmentation give essentially the same expression for the minimum mass ($M_{_{\rm MIN2}} \sim M_{_{\rm MIN3}}$), they give very different expressions for the Jeans mass ($M_{_{\rm J2}} \neq M_{_{\rm J3}}$), and there are other important differences. In particular, 2-D one-shot fragmentation bypasses all the problems associated with 3-D hierarchical fragmentation. There is no backwarming because there are no other local fragments filling the part of the celestial sphere into which a fragment radiates (i.e. perpendicular to the layer). More importantly, there is little likelihood of fragments merging, since fragments with $M_{_{\rm F2}} \sim M_{_{\rm J2}}$ condense out faster than any other structures in a layer (whereas fragments in a 3-D medium with $M_{_{\rm F3}} \sim M_{_{\rm J3}}$ condense out slower than all larger structures in that medium; e.g. {\it Larson}, 1985). Finally, because condensation in a layer is so fast, growth by accretion is limited.

{\it Boyd and Whitworth} (2004) have analysed the radiative cooling of a fragmenting layer, taking into account {\it not only} the $PdV$ heating of a condensing fragment, {\it but also} the on-going accretion (as matter continues to flow into the layer) {\it and} the energy dissipated in the accretion shock. They find that for $T \sim 10\,{\rm K}$, the minimum mass is $0.0027\,{\rm M}_{_\odot}$. This fragment starts with mass $\sim 0.0011\,{\rm M}_{_\odot}$, but continues to grow by accretion as it condenses out.

\subsection{\textbf{Fragmentation of a circumstellar disc}}\label{SEC:MIND}

Another scenario which may be more relevant to contemporary star formation than 3-D hierarchical fragmentation is fragmentation of a circumstellar disc. There are three critical issues here. (i) Under what circumstances does a disc fragment gravitationally? (ii) Can the resulting fragments cool fast enough to condense out? And (iii) can the resulting fragments lose angular momentum fast enough to condense out? The last two issues are critical because, if a fragment cannot cool and lose angular momentum fast enough, it is likely to bounce and be sheared apart. For simplicity we consider an equilibrium disc.

(i) The condition for an isolated disc to fragment gravitationally is that the surface density, $\Sigma$, be sufficiently large, 
\begin{equation} \label{EQN:TOOMRE}
\Sigma \;\ga\; \Sigma_{_{\rm T}} \;\simeq\; \frac{a\,\epsilon}{\pi\,G},
\end{equation}
where $a$ is the local sound speed and $\epsilon$ is the local epicyclic frequency ({\it Toomre}, 1964).

Condition (\ref{EQN:TOOMRE}) is also the condition for spiral modes to develop in the disc, and these will have the effect of redistributing angular momentum. As a result, the inner parts of the disc may simply accrete onto the central primary star and the outer parts may disperse {\it without fragmenting} ({\em Laughlin and Bodenheimer}, 1994; {\em Nelson et al.}, 1998). Thus, if fragmentation is to occur, it must occur on a dynamical timescale. 

(ii) The condition for a fragment to cool fast enough to condense out is therefore that the fragment can radiate away, on a dynamical timescale, the thermal energy being delivered by compression. Gammie (2001) has shown that for a Keplerian disc, this condition can be written as a constraint on the cooling time,
\begin{eqnarray} \label{EQN:GAMMIE}
t_{_{\rm COOL}} \;\la\; \frac{3}{\Omega} \,,
\end{eqnarray}
where $\Omega$ is the local orbital angular speed.

We assume that, as a disc forms, $\Sigma$ increases sufficiently slowly that it does not greatly exceed $\Sigma_{_{\rm T}}$ when the disc becomes unstable. It then follows that the radius, growth time and mass of the fastest growing fragment are
\begin{eqnarray}
R_{_{\rm FD}} & \simeq & \frac{a}{\epsilon} \;; \\
t_{_{\rm FD}} & \simeq & \frac{1}{\epsilon} \;; \\
M_{_{\rm FD}} & \simeq & \frac{a^3}{G\epsilon}.
\end{eqnarray}
The compressional heating rate for a fragment is thus 
\begin{equation}
{\cal H} = P\,\frac{dV}{dt} \simeq \frac{3 M_{_{\rm FD}} a^2}{2 t_{_{\rm FD}}} \simeq \frac{3 a^5}{2 G};
\end{equation}
and the radiative cooling rate is
\begin{equation}
{\cal L} \simeq \frac{2\pi R_{_{\rm FD}}^2\sigma_{_{\rm SB}}T^4}{\left(\bar{\tau}_{_{\rm R}}(T)+\bar{\tau}_{_{\rm P}}^{\;-1}(T)\right)} \simeq \frac{4\pi^6\bar{m}^4a^{10}/15c^2h^3\epsilon^2} {\left(\bar{\tau}_{_{\rm R}}(T)+\bar{\tau}_{_{\rm P}}^{\;-1}(T)\right)},
\end{equation}
where now $\bar{\tau}_{_{\rm R}}(T) \simeq \bar{\tau}_{_{\rm P}}(T) \simeq a \epsilon \bar{\kappa}(T) / \pi G$. Consequently the requirement that ${\cal L} \ga {\cal H}$ reduces to
\begin{equation} \label{EQN:GAMMIE1}
\frac{\epsilon^2}{a^5} \la \frac{4\pi^6G\bar{m}^4/45c^2h^3}{\left(\bar{\tau}_{_{\rm R}}(T)+\bar{\tau}_{_{\rm P}}^{\;-1}(T)\right)}.
\end{equation}

To illustrate the discussion we consider the specific case of a Keplerian disc around a Sun-like star, with
\begin{eqnarray}
\epsilon(D) & \simeq & 2 \times 10^{-7}\,{\rm s}^{-1} \left(\frac{M_{\star}}{{\rm M}_{_\odot}}\right)^{1/2} \left(\frac{D}{\rm AU}\right)^{-3/2},\;\;\;\;\;\;\;\; \\
T(D) & \simeq & 300\,{\rm K} \left(\frac{L_{\star}}{{\rm L}_{_\odot}}\right)^{1/4} \left(\frac{D}{\rm AU}\right)^{-1/2}, 
\end{eqnarray}
and hence $a(D) \simeq 1\,{\rm km}\,{\rm s}^{-1} \left(L_{\star}/{\rm L}_{_\odot}\right)^{1/8} \left(D/{\rm AU}\right)^{-1/4}$, where $D$ is distance from the Sun-like star. The fastest growing fragment then has mass 
\begin{equation}
M_{_{\rm F\,D}} \simeq 3 \times 10^{-5}\,{\rm M}_{_\odot} \left(\frac{M_{\star}}{{\rm M}_{_\odot}}\right)^{-1/2} \left(\frac{L_{\star}}{{\rm L}_{_\odot}}\right)^{3/8} \left(\frac{D}{\rm AU}\right)^{3/4},
\end{equation} 
and Condition (\ref{EQN:GAMMIE1}) is only satisfied for
\begin{eqnarray} \label{EQN:DMIND}
D & \ga & 90\,{\rm AU} \left(\frac{M_{\star}}{{\rm M}_{_\odot}}\right)^{3/7} \left(\frac{L_{\star}}{{\rm L}_{_\odot}}\right)^{-1/7}, \\ \label{EQN:MMIND}
M_{_{\rm FD}} & \ga & 0.003\,{\rm M}_{_\odot} \left(\frac{M_{\star}}{{\rm M}_{_\odot}}\right)^{-5/28} \left(\frac{L_{\star}}{{\rm L}_{_\odot}}\right)^{15/56}.\hspace{0.8cm}
\end{eqnarray}

(iii) Angular momentum is removed from a condensing fragment by gravitational torques, in the same way that angular momentum is redistributed in the disc as a whole when $\Sigma \ga \Sigma_{_{\rm T}}$. This is a non-linear and stochastic process, and there is no analytic estimate of the rate at which it occurs. Therefore the condition for a fragment to lose angular momentum fast enough to condense out has to be evaluated by numerical simulation. The chapter by {\em Durisen et al.} deals with this problem.

\subsection{\textbf{Non-linear thermodynamics}}

Impulsive triggers which produce rapid compression will always help to amplify self-gravity, because the freefall time varies as $t_{_{\rm FF}} \propto \rho^{-1/2}$. However, there is only a very restricted temperature range within which rapid compression will help an optically thick proto-fragment to cool more rapidly (and thereby avoid the likelihood of its bouncing and being sheared apart). Suppose that the Rosseland-mean opacity is given by $\bar{\kappa}_{_{\rm R}} \propto \rho^\alpha T^\beta$, and that the gas has a ratio of specific heats $\gamma$.

Then if a fragment condenses out quasistatically, its cooling time varies as $t_{_{\rm COOL}} \propto \rho^{\alpha + (1+\beta)/3}$, so $t_{_{\rm COOL}}$ only decreases as fast as $t_{_{\rm FF}}$, with increasing density, if $\beta < -(6\alpha + 5)/2$. This condition is only likely to be satisfied in the temperature range where refractory dust sublimates ($1500\;{\rm to}\;3000\,{\rm K}$).

If instead the fragment is compressed impulsively -- and therefore adiabatically -- its cooling time varies as $t_{_{\rm COOL}} \propto \rho^{\alpha + 4/3 + (\beta - 3)(\gamma -1)}$, so now $t_{_{\rm COOL}}$ only decreases as fast as $t_{_{\rm FF}}$, with increasing density, if $\beta < 3 - (6\alpha + 11)/6(\gamma - 1)$. For $T \la 100\,{\rm K}$, we have $\gamma \simeq 5/3$ and so enhanced cooling requires $\beta \la -1/4$, which is unlikely. For $100\,{\rm K} \la T \la 1000\,{\rm K}$, $\gamma \simeq 7/5$  and enhanced cooling requires $\beta \la -19/12$, which is even less likely. For $1,000\,{\rm K} \la T \la 3,000\,{\rm K}$, H$_{_2}$ dissociation gives $\gamma \sim 1.1$  and enhanced cooling requires $\beta \la -15$, which may occur during sublimation, but then necessarily only over a small temperature range. For $3,000\,{\rm K} \la T \la 10,000\,{\rm K}$, $\gamma \sim 5/3$ and H$^{^-}$ opacity gives $\alpha \sim 1/2$ and $\beta \sim 4$, so $t_{_{\rm COOL}} \propto \rho^{7/2}$. At even higher temperatures where Kramers opacity dominates $t_{_{\rm COOL}} \propto \rho^{-2}$, and where electron scattering opacity dominates $t_{_{\rm COOL}} \propto \rho^{-2/3}$, but by this stage a fragment is very opaque and very strongly bound.

\subsection{\textbf{Forming close BD-BD binaries}}

It is possible that there is a secondary fragmentation regime at $T \sim 2000\,{\rm K}$, due to the softening of the equation of state caused by H$_{_2}$ dissociation, and the enhanced cooling which occurs in the opacity gap between dust sublimation and H$^{^-}$ opacity. A spherical cloud of mass $M$ in hydrostatic equilibrium at this temperature has radius 
\begin{eqnarray} \label{EQN:BDBDBIN}
R & \sim & 4\,{\rm AU}\,\left( \frac{M}{0.1\,{\rm M}_{_\odot}} \right) \,.
\end{eqnarray}
If the cloud fragments into a binary and virialises, the binary should have a separation of the same order. Eqn. (\ref{EQN:BDBDBIN}) is actually a good mean fit to the values plotted on Fig. 6 of the chapter by {\em Burgasser et al.}, suggesting that BD-BD binaries may be produced by secondary fragmentation facilitated by H$_{_2}$ dissociation and/or the opacity gap.

\subsection{\textbf{Forming close BD-BD binaries in discs}}

{\em Burgasser et al.} (2005) have noted that -- modulo the small-number statistics involved -- a brown dwarf in a wide orbit ($\ga 200\,{\rm AU}$) about a Sun-like star is apparently more likely to be in a close BD-BD binary system ($\la 20\,{\rm AU}$) than a brown dwarf in the field. If this trend is confirmed, it suggests that BD-BD binaries are formed in discs, and then may be ejected. For a close BD-BD binary in a wide orbit around a Sun-like star, the internal binding energy of the BD-BD binary is typically comparable with or larger than the binding energy of the BD-BD binary to the Sun-like star, and therefore some BD-BD binaries should be able to survive ejection.

Moreover, we know, from the Toomre criterion, that for fragments condensing out of discs, losing angular momentum is a critical issue. The smaller a condensation becomes, the slower the rate at which angular momentum can be lost by gravitational torques. Therefore such a condensation may be strongly disposed to binary fragmentation when its thermal support is weakened at $T \sim 2000\,{\rm K}$. Since only a subset of brown dwarfs is in close BD-BD binaries, this suggestion does not require that all brown dwarfs are formed in discs.

\section{\textbf{FORMING VERY LOW-MASS PRESTELLAR CORES IN TURBULENT CLOUDS}} \label{SEC:TURB}%

A number of surveys ({\em Testi and Sargent}, 1998; {\em Peng et al.}, 1998; {\em Motte et al.}, 1998; {\em Motte et al.}, 2001; {\em Motte and Andr\'e}, 2001; {\em Johnstone et al.}, 2000; {\em Sandell and Knee}, 2001; {\em Onishi et al.}, 2002; {\em Reid and Wilson}, 2005) have noted the close similarity between the core mass function (CMF) and the stellar initial mass function (IMF). Cores with brown dwarf masses have been observed in several of these surveys, but since these low-mass cores are usually below the completeness limit, their statistics are unreliable. However, the similarity between the CMF and the IMF suggests that each core gives rise to a similar low number of stars, whose final masses are heavily influenced by the mass of the core. Indeed, protostars are generally observed inside relatively well defined envelopes, as implied by the definition of Class 0 and Class I objects ({\em Andr\'e et al.}, 1993, 2000; {\em Tachihara et al.}, 2002). This is not to say that the final masses of stars are the same as the masses of the prestellar cores in which they form. Protostars are associated with outflows, which can remove a significant fraction of the mass of the protostellar envelope, and cores may split into more than one star (see also {\em Hosking and Whitworth}, 2004; {\em Stamatellos et al.}, 2005). However, low mass cores appear typically to harbour only one or two protostars  (e.g. {\em Tachihara et al.}, 2002), and the overall binary statistics are inconsistent with a larger number of stars being formed from each core ({\em Goodwin and Kroupa}, 2005).


\subsection{Core structure and low mass core formation}

Observed prestellar cores generally have Bonnor-Ebert (BE) like density profiles, with relatively sharp outer edges ({\em Bacmann et al.}, 2000; {\em Motte and Andr\'e}, 2001; {\em Kirk et al.}, 2005). Starless cores have flat BE-like profiles, while cores with detected Class 0 or Class I protostars have more centrally peaked density profiles, presumably as a result of their deeper potential wells. Since well defined cores with sharp edges define finite mass reservoirs for the stars that form inside them, the core observations reinforce the notion that the masses of stars are, at least statistically, strongly influenced by the masses of the cores within which they form ({\em Padoan and Nordlund}, 2002, 2004).

It is therefore important to consider whether it is possible to form brown dwarfs directly, from correspondingly low-mass cores. The standard argument against brown dwarfs being formed directly is that the density needs to be very high for a fragment with brown dwarf mass to collapse gravitationally. If one considers low-amplitude perturbations on top of a typical mean density of $10^4$ to $10^5\,{\rm cm}^{-3}$, he will conclude that only fluctuations of order the Jeans mass ($1\;{\rm to}\;3\,{\rm M}_{_\odot}$; Eqn. \ref{EQN:MJEANS3}) are able to collapse in a typical molecular cloud.  However, many cores actually have very high density contrast relative to their surroundings, and so it is inappropriate to use the Jeans mass at the mean density as an estimate of the resulting protostellar mass.

It is also important to realise that cores having sufficiently high density to form brown dwarfs directly need not -- indeed must not -- be very common, and therefore we can appeal to exceptional circumstances to generate them. As a measure of how exceptional the circumstances must be, consider the IMF (in the form $d{\cal N}/d\ln N$). At high masses, the IMF falls off with the {\em Salpeter} (1955) exponent $-1.35$, and if this power-law IMF continued unbroken down to the brown-dwarf regime, $10^{4\times 1.35} \simeq 250,000$ brown dwarfs with $M \sim 0.01\,{\rm M}_{_\odot}$ would be formed for every massive star with $M \sim 100\,{\rm M}_{_\odot}$. Instead, the IMF peaks near $\sim 0.3\,{\rm M}_{_\odot}$ and falls off at low masses; adopting the {\em Chabrier} (2003) IMF for the Milky Way, we find that the number of brown dwarfs formed with $M \sim 0.01\,{\rm M}_{_\odot}$ is actually about equal to the number of massive stars formed with $M \sim 100\,{\rm M}_{_\odot}$, as predicted theoretically by {\em Zinnecker} (1984).

In order for a high-density core to form, its mass must be concentrated into a small volume.  If ${\bf u}$ is the fluid velocity, accumulation of mass is measured by the quantity $-\nabla\cdot{\bf u} = D \ln\rho/Dt$ (the comoving time derivative of log density), and so individual cores form at local maxima of $-\nabla\cdot{\bf u}$.

The flow towards a convergence point is generally supersonic, and so stand-off shocks develop, separating the unshocked upstream gas from the shocked and nearly stagnant downstream gas. The stand-off shocks correspond to density jumps $\sim{\cal M}^2$ if the shocked gas is dominated by gas pressure; and $\sim {\cal M}_{_{\rm A}}$ (the Alfv\'enic Mach number) if the shocked gas is dominated by magnetic pressure. As matter accumulates around the convergence point, a deepening gravitational potential well develops, and the accumulated and stagnant gas forms a growing Bonnor-Ebert like core, stabilised by the external ram pressure, $\rho |{\bf u}|^2$.

If the density at the centre of the core increases to more than about 14 times the boundary density (just inside the shock surface), the core becomes gravitationally unstable and collapses. However, even before it becomes unstable, such a core has -- by virtue of the shock jump at its boundary -- a total density contrast (between its centre and the surrounding inflowing unshocked gas) that can greatly exceed a factor of 14. Moreover, if the inflowing gas runs out before the core has become sufficiently massive and dense to collapse, then the core will expand and disperse due to the decrease in ram pressure at its boundary.

In the last column of Table 3 in {\em Bacmann et al.} (2000), at least 8 of the 9 cores listed have density contrasts exceeding 14, but no strong indication of collapse. {\em Bacmann et al.} conclude that the structures are best fitted with Bonnor-Ebert spheres, but since they cannot explain how the cores can be stable at these high density contrasts, they go on to discuss other density profiles and models, which do not fit the data as well as BE-spheres. Similar core profiles have been observed in many other surveys. There is thus direct observational evidence for the creation of prestellar cores with large density contrasts between their centres and the surrounding medium, as expected when cores are produced as stagnant structures by supersonic flows. Indeed, cores formed in numerical simulations of molecular cloud turbulence have internal velocity dispersions and rotation velocities that are entirely consistent with those of observed cores (cf. Figs. 8 and 9 in {\em Nordlund and Padoan}, 2003).

With this scenario in mind, we now investigate semi-quantitatively the range of densities that is possible, and the circumstances under which the densities attained would be high enough to lead to the collapse of brown dwarf mass cores. Any particular converging flow may be broadly characterised by three parameters: the up-stream density $\rho_0$, the Mach number ${\cal M}$ of the up-stream flow, and the degree of focussing of the up-stream flow towards a 3-D convergence point.  As a measure of the latter we take the ratio $f$ of the surface area that encloses the up-stream flow to the surface area of the stand-off shock that surrounds the central BE-like structure. Since the up-stream flow is supersonic, its motion is essentially inertial (i.e. constant velocity), until it encounters the stand-off shock.  Thus, by mass conservation, its density increases from $\rho_0$ to $f \rho_0$. At the stand-off shock the density increases by an additional factor ${\cal M}^2$ for hydro-shocks (or ${\cal M}_{_{\rm A}}$ for MHD-shocks).  Finally, from just inside the shock to the centre of the BE-like core the density increases by a further factor $\la 14\;$ ($\sim 14$, in the marginally stable case). The total density increase from the up-stream source to the core centre is $\sim 14 f {\cal M}^2$.  This factor has the right dependence to account for the formation of brown dwarf mass cores, in that it can be arbitrarily large, albeit with decreasing probability.

\begin{figure}[h]
\epsscale{1.0}
\plotone{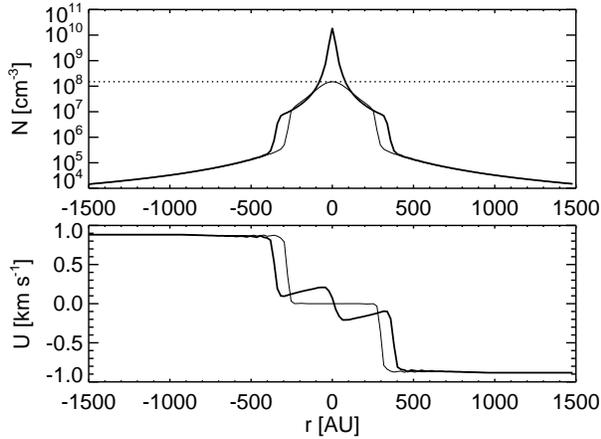}
\caption{\small Density- and velocity-profiles, through a core forming from a spherically symmetric convergent flow. The thin line represents an early stage when the $\sim 0.015\,{\rm M}_{_\odot}$ core approximates to a stable and stagnant BE-like configuration; the boundary shock is at $\sim 300\,{\rm AU}$. The thick line represents a later stage after the core has become unstable and started to collapse; the boundary shock is now at $\sim 350\,{\rm AU}$, because the core is more massive ($\sim 0.025\,{\rm M}_{_\odot}$). Outside the shock is the convergent inertial flow. Note the very large density contrast, even between the gas at the edge of the core and the ambient medium at $\sim 1500\,{\rm AU}$.}  
\end{figure}

As an example, consider the formation of a core with temperature $\sim 10\,{\rm K}$ and central density $\sim 10^7\,{\rm cm}^{-3}$, sufficiently high to form a star with $M \sim 0.1\,{\rm M}_{_\odot}$. If the average cloud density is $\sim 10^4\,{\rm cm}^{-3}$ and the Mach number is 5, $f$ only needs to be $\sim 3$; i.e., the up-stream flow needs to be focussed onto a three times smaller stand-off shock surface. This must be rather common, and indeed at $M \sim 0.1\,{\rm M}_{_\odot}$ the IMF has hardly dropped below its maximum.

Repeating this estimate for a core with  $M \sim 0.01\,{\rm M}_{_\odot}$ one finds that the flow needs to be much better focussed. The stand-off shock surface now needs to be $\sim 300$ times smaller than the up-stream source area, and so the linear size of the core needs to be $\sim 17$ times smaller than the scale of the up-stream flow. This does require a rather exceptional focussing of the up-stream flow, but then brown dwarfs with $M \sim 0.01\,{\rm M}_{_\odot}$ actually are very rare.  The fraction of brown dwarfs that form in this manner can only be ascertained quantitatively by performing very high resolution numerical simulations.

Fig. 1 illustrates an example intermediate between these two extremes, i.e. a core with final mass $\sim 0.03\,{\rm M}_{_\odot}$.

\subsection{\textbf{Envelope break-out and continued accretion}}

The sharp density transitions surrounding accreting cores produce more well defined cores than would exist in a subsonic medium. Nevertheless there will be some continued accretion after a collapsed object is formed, until the upstream supply of gas is exhausted. Except towards the end of the process, such cores correspond to Class 0 objects, which by definition have not yet accreted half their final mass ({\em Andr\'e et al.}, 2000).

Class I and II objects are observed to accrete from low mass envelopes ({\em Motte and Andr\'e}, 2001), and have much reduced accretion rates, $\la 10^{-6}\,{\rm M}_{_\odot}\,{\rm yr}^{-1}$.  Once the natal envelope is consumed there is no longer a strong coupling between the collapsed object (moving under the N-body influence of neighbouring objects and gas) and the surrounding medium, and one expects the collapsed object to pick up speed relative to the surrounding gas, and to then accrete in a manner similar to Bondi-Hoyle accretion ({\em Padoan et al.}, 2005; {\em Krumholz et al.}, 2005b). In a turbulent medium the problem is more complicated ({\em Krumholz et al.}, 2005a) , but the scaling is more or less as for Bondi-Hoyle accretion, with a prefactor $\Phi \sim 1\;{\rm to}\;5$ that accounts for the complications. Quantitative estimates indicate that the accretion after break-out from the natal core is insignificant.

\subsection{\textbf{The viability of turbulent fragmentation}}

Although turbulent fragmentation generates a mass function for prestellar cores, matching broadly the observed stellar IMF, there are two caveats which should be born in mind when considering the formation of brown dwarfs.

First, in turbulent fragmentation ({\em Padoan and Nordlund}, 2002), brown dwarfs are produced by a small subset of low-mass cores, i.e. those which are sufficiently dense to collapse gravitationally. At its low-mass end, the overall (i.e. fully sampled) CMF should be dominated by a much larger number of transient (i.e. non-prestellar) cores. Moreover, many of these transient cores will be only slightly less dense than the ones that spawn brown dwarfs. They will therefore be detectable in submillimetre continuum observations, and they should also be somewhat longer lived. Thus, even allowing for selection effects, the observed CMF should fall much less steeply with decreasing mass than the stellar IMF. Recent estimates of the CMF ({\em Nutter \& Ward-Thompson}, in prep.), which probe to lower masses by using longer exposures, actually suggest the opposite, but completeness remains a concern (e.g. {\em Kirk et al.}, 2006).

Second, turbulent fragmentation predicts a ratio of brown dwarfs to H-burning stars, ${\cal R}$, which is exponentially sensitive to the Alfv\'enic Mach number on the largest scales (${\cal M}_{_{\rm A}}$) and to the mean cloud density ($n$). Regions with smaller values of ${\cal M}_{_{\rm A}}$ and/or $n$ should generate stellar populations with significantly fewer brown dwarfs, and regions with larger values should generate stellar populations with significantly more brown dwarfs. However, in nature ${\cal R}$ appears to vary little over a wide range of local star-forming environments (see chapter by {\em Luhman et al.}).

This problem can be overcome if nature selects a narrow range of ${\cal M}_{_{\rm A}}$ and $n$. For example, Whitworth (2005) has noted that contemporary star formation may only proceed rapidly if the gas couples thermally to the dust (so that it can avail itself of broadband -- as distinct from molecular-line -- cooling). This requires that the ram pressure in shocks producing prestellar cores exceeds a critical value, which converts into the constraint $nT{\cal M}_{_{\rm A}}^2 \ga P_{_{\rm CRIT}} \sim 10^5\,{\rm cm}^{-3}\,{\rm K}$. This constraint may help to select the combinations of ${\cal M}_{_{\rm A}}$ and $n$ which reproduce the observed ratio of brown dwarfs to H-burning stars.

\section{\textbf{COLLAPSE AND FRAGMENTATION OF LARGE PRESTELLAR CORES}} \label{SEC:FRAG}%

Whilst a very low-mass prestellar core ($\la 0.1\,{\rm M}_{_\odot}$) must collapse to form either a single brown dwarf or a multiple brown dwarf system, larger prestellar cores ($\ga 1\,{\rm M}_{_\odot}$) are expected to form clusters of stars having a range of masses. We can identify five mechanisms which may play a role in determining the final stellar masses. (i) During the approximately isothermal initial collapse phase, as the pressure becomes increasingly unimportant and the collapse approaches freefall, self-gravity will amplify any existing density substructure. (ii) Then, when the density reaches $n_{_{{\rm H}_2}} \sim 10^{11}\,{\rm cm}^{-3}$, the gas becomes optically thick and switches rather suddenly from approximate isothermality to approximate adiabaticity. At this juncture, a network of shock waves develops to slow the collapse down, and non-linear interactions between these shock waves produce and amplify further substructure (sheets, filaments and isolated prestellar cores). (iii) Some of these structures will have sufficient angular momentum to form discs, and these may then fragment due to rotational instability (see \S \ref{SEC:DISC}). Finally, once a protostellar embryo (i.e. an object which is sufficiently well bound to be treated dynamically as a single entity) has condensed out of a fragment, its subsequent evolution and final mass will be determined by (iv) competitive accretion and (v) dynamical interaction (see \S \ref{SEC:EJEC}). Here we concentrate on mechanisms (i) and (ii), since these are the ones which distinguish the evolution of a high-mass core from the low-mass cores considered in \S {\ref{SEC:TURB}.

High-mass prestellar cores ($\ga 10\,{\rm M}_{_\odot}$) invariably display non-linear internal density structure, and at the typical density ($n_{{\rm H}_2} \sim 10^5\,{\rm cm}^{-3}$) and temperature ($T \sim 10\,{\rm K}$) in a prestellar core, the Jeans mass is $M_{_{\rm J3}} \sim 0.8\,{\rm M}_{_\odot}$ (Eqn. \ref{EQN:MJEANS3}). Therefore, in the absence of a significant magnetic field, they are very likely to fragment during collapse.

Even the smallest cores are usually far from spherical, and have been modelled as being either prolate or oblate ({\it Myers et al.}, 1991; {\it Ryden}, 1996; {\it Jones et al.}, 2001; {\it Goodwin et al.}, 2002; {\it Curry}, 2002; {\it Myers}, 2005) with prolate models being favoured statistically, although in reality cores are probably triaxial -- or even more complicated -- in their full three dimensional structure ({\it Jones et al.}, 2001; {\it Goodwin et al.}, 2002). Gravity works to enhance anisotropies in collapsing objects ({\it Lin et al.}, 1965) with collapse occurring fastest along the shortest axis to form sheets and filaments, which then subsequently fragment ({\it Bastien}, 1983; {\it Inutsuka and Miyama}, 1992). Thus, it is unsurprising that hydrodynamical simulations of both oblate and prolate cores are prone to fragmentation (e.g. {\it Bastien et al.}, 1991; {\it Boss}, 1996).

Prestellar cores also tend to have complex internal velocity fields ({\it Larson}, 1981; {\it Myers and Benson}, 1983; {\it Arquilla and Goldsmith}, 1985), but since prestellar cores can only be observed from a single direction, the interpretation of these velocities is difficult. If cores are assumed to be in solid-body rotation, the ratio of rotational to gravitational energy is typically $\beta \equiv {\cal R}/|{\Omega}| \sim 0.03$, with some cases as high as $\beta \sim 0.1$ ({\it Goodman et al.}, 1993; {\it Barranco and Goodman}, 1998). However, the observed velocities are more likely to be turbulent in nature, i.e. less well ordered than solid-body rotation ({\it Myers and Gammie}, 1999; {\it Burkert and Bodenheimer}, 2000).  Quite low levels of turbulence (e.g. {\it Goodwin et al.} 2004a), and/or global rotation (e.g. {\it Boss}, 1986; {\it Bonnell and Bate}, 1994a; {\it Hennebelle et al.}, 2004; {\it Cha and Whitworth}, 2004) are sufficient to make a collapsing core fragment into a small ensemble of protostellar embryos.

Unfortunately, collapse and fragmentation can only be explored by means of numerical simulations, and there is a huge and poorly constrained range of admissible initial conditions, which makes the extraction of robust theorems very hard (see e.g. {\it Hennebelle et al.}, 2004). Moreover almost all simulations to date use barotropic equations of state (i.e. $P = P(\rho)$). These barotropic equations of state are designed to mimic the expected thermal behaviour of protostellar gas, but they do not capture the thermal inertia effects which become important at the juncture when the gas starts to heat up due to adiabatic compression, and it appears ({\it Boss et al.}, 2000) that these thermal inertia effects play a critical, deterministic role in gravitational fragmentation. Proper treatment of the energy equation and the associated radiative transport (e.g. {\em Whitehouse and Bate}, 2006) is needed to make these simulations more realistic.

\section{\textbf{DISC FRAGMENTATION}} \label{SEC:DISC}%

We organise our discussion of disc fragmentation under the headings (i) isolated, relaxed discs, (ii) unrelaxed discs, and (iii) interacting discs.

\subsection{\textbf{Isolated relaxed discs}}

The dynamical fragmentation of isolated relaxed discs is discussed in detail in the chapter by {\em Durisen et al}. Although the emphasis there is on the genesis of planets, the same issues pertain to the formation of brown dwarfs, viz. Under what circumstances do discs become unstable against fragmentation? Is Eqn. (\ref{EQN:TOOMRE}) a sufficient condition for gravitational instability (in which case, what is the precise value of $\Sigma_{_{\rm T}}$) or is it also necessary for $\Sigma$ to increase rapidly (in order to avoid the disc simply being accreted and dispersed by torques due to spiral density waves)? Does Eqn. (\ref{EQN:GAMMIE}) determine whether fragments can cool fast enough to condense out? What role is played by thermodynamic effects like H$_{_2}$ dissociation, dust sublimation, and convection? What are the properties of dust in discs, and how well mixed are the gas and dust? Does the survival of a fragment depend on its ability to lose angular momentum rapidly? Some of these issues are only beginning to be investigated.

{\it Rice et al.} (2003) have corroborated Eqn. (\ref{EQN:GAMMIE}) numerically by performing SPH simulations of disc fragmentation with a parameterised cooling law of the form $du/dt = - u \Omega / \beta$ (where $u$ is the specific internal energy and $\Omega$ the orbital angular speed). Endemic fragmentation occurs with $\beta = 3$ but not with $\beta = 5$. However, because this cooling law results in indefinite cooling, whereas there is a limit to the supply of rotational energy which can be tapped through shock heating, by the time fragmentation occurs the temperatures have dropped to rather low values, which may be hard to realise in nature.

{\it Rafikov} (2005) argues that the opacity of gravitationally unstable discs is so high, and the cooling times are therefore so long, that fragments can only condense out on a dynamical timescale in the outer, cooler regions of massive discs. (Although the treatment is somewhat different, his conclusion resonates with the simple analysis we have presented in \S 3.3 and carries the same caveats). 

{\it Johnson and Gammie} (2003) point out that the effects of opacity may be mitigated in temperature regimes where dust sublimates and hence the opacity decreases abruptly with increasing temperature. However, dust sublimation effects are confined to the temperature range $T \ga 200\,{\rm K}$, and the most critical effects occur at $T \ga 2000\,{\rm K}$, so they are probably only relevant to fragmentation close to the central star ($D \la 10\,{\rm AU}$).

{\it Cai et al.} (2005) report simulations of disc evolution taking radiation transport into account, and conclude that -- even at low metallicities -- the opacity in the hot inner parts of a disc is too high to allow fragmentation on a dynamical timescale, in agreement with {\it Rafikov} (2005).

In contrast, {\it Boss} (2004) presents simulations of disc evolution taking radiation transport into account and concludes that fragments of planetary mass do condense out -- or at least that gravitationally bound fragments form which would subsequently condense out. He argues that the proto-fragments in his simulations bypass the effects of high opacity by transporting energy convectively. However, it is not clear how convection can cool a fragment which is condensing out on a dynamical timescale. The coherent small-scale motions he attributes to convection may actually be manifestations of a fragment which is unable to cool and is bouncing -- or will soon bounce -- prior to being sheared apart. This needs to be investigated further, but the numerical complexities are considerable.

\subsection{\textbf{Unrelaxed discs}}

In numerical simulations of the collapse and fragmentation of intermediate- and high-mass prestellar cores, the first single protostars to form usually quickly acquire massive circumstellar discs, and secondary protostars then condense out of these discs. This pattern is common for rotating cores (e.g. {\em Bonnell}, 1994; {\em Bonnell and Bate}, 1994a, 1994b; {\em Turner et al.}, 1995; {\em Whitworth et al.}, 1995; {\it Burkert and Bodenheimer}, 1996; {\em Burkert et al.}, 1997), for turbulent cores (e.g. {\em Bate et al.}, 2002b, 2003; {\it Goodwin et al.}, 2004b) and for cores which are subjected to a sudden increase in external pressure (e.g. {\it Hennebelle et al.}, 2004). The discs thus formed fragment before they have time to relax to an equilibrium state. Indeed, the material accreting onto the disc is often quite lumpy, and this helps to seed fragmentation. Under this circumstance, gravitational fragmentation is more likely simply because proto-fragments are launched directly into the non-linear regime of gravitational instability, rather than having first to grow through the linear phase.

Even so, a proto-fragment still has to be able to cool and lose angular momentum on a dynamical timescale, if it is to condense out, rather than bouncing and then being sheared apart. Fragmentation of unrelaxed discs can only be studied by means of numerical simulations, and to date no simulations of the process have been performed which treat properly the energy equation and the associated transport of radiation. Since young protostars have relatively high luminosities, an important consideration will be irradiation of the disc by the primary protostar at its centre and any other nearby protostars.

\subsection{\textbf{Interacting discs}}

Another way in which disc fragmentation can be triggered is by an impulsive interaction with another disc, or with a naked star. In the dense protocluster environment where most stars are presumed to form, such interactions must be quite frequent, since many very young protostellar discs have diameters $\ga 300\,{\rm AU}$ and the mean separation between neighbouring protostars in a typical cluster is $\la 3000\,{\rm AU}$. Indeed, $\sim 50\%$ of solar-type stars end up in binary systems with semi-major axes $a \la 1000\,{\rm AU}$, so the notion of a disc evolving in the gravitational field of a single, isolated protostar is probably rather artificial. 

{\it Boffin et al.} (1998) and {\it Watkins et al.} (1998a, 1998b) have simulated parabolic interactions between two protostellar discs, and between a single protostellar disc and a naked protostar. All possible mutual orientations of spin and orbit are sampled, and the gas is assumed to behave isothermally, which is probably a reasonable assumption, since the discs are large (initial radius $1000\,{\rm AU}$) and most of the secondary protostars form at large distances (periastra $\ga 100\,{\rm AU}$). The critical parameter turns out to be the effective shear viscosity in the disc. If the Shakura-Sunyaev parameter is low, $\alpha_{_{\rm SS}} \sim 10^{-3}$, most of the secondary protostars have masses in the range $0.001\,{\rm M}_{_\odot}$ to $0.01\,{\rm M}_{_\odot}$. Conversely, if $\alpha_{_{\rm SS}}$ is larger, $\alpha_{_{\rm SS}} \sim 10^{-2}$, most of the secondary protostars have masses in the range $0.01\,{\rm M}_{_\odot}$ to $0.1\,{\rm M}_{_\odot}$. The formation of low-mass companions is most efficient for interactions in which the orbital and spin angular momenta are aligned; on average 2.4 low-mass companions are formed per interaction in this case. If the orbital and spin angular momenta are randomly orientated, then on average 1.2 companions are formed per interaction. It is important that such simulations be repeated, with a proper treatment of the energy equation and the associated energy transport, to check their fidelity, and to establish whether low-mass companions can form at closer periastra.

\section{\textbf{PREMATURE EJECTION OF PROTOSTELLAR EMBRYOS}} \label{SEC:EJEC}%

The scenario where brown dwarfs form by premature ejection is closely linked to - but ultimately independent of -- the notion of competitive accretion. In the ejection hypothesis, brown dwarfs are simply protostellar embryos which get separated from their reservoir of accretable material at an early stage, and so in the context of brown dwarf formation it is irrelevant whether there are other stars competing for this same material. The importance of competitive accretion for intermediate and high-mass stars is argued in the chapter by {\em Bonnell et al.} The contentious issue is whether protostellar embryos -- i.e. the first, very low-mass ($\sim 0.003\,{\rm M}_{_\odot}$) high-density star-like objects -- exist for long enough to do much competing. At one extreme, it is argued that a protostellar embryo forms following the non-homologous collapse of a much larger gravitationally unstable core, and therefore it accretes mainly from its own co-moving placenta. At the other extreme, it is argued that a protostellar embryo quickly becomes sufficiently decoupled from the ambient gas that it can roam around competing with other embryos for the same reservoir of accretable gas. For the purpose of this section, we assume that nature cleaves to the second extreme.

\subsection{Competitive accretion in gas-rich proto-clusters}

Once an ensemble of protostellar embryos has formed, still deeply embedded in its parental prestellar core and/or parental disc, the individual embryos evolve by accreting gas, and by interacting dynamically with one another. The accretion histories of individual embryos differ due to their varying circumstances, leading to a spectrum of protostellar masses, extending from high masses down to below the H-burning limit. Those which spend a long time moving slowly through the dense gas near the centre of the core can grow to high mass.  Conversely, those that spend most of their time moving rapidly through the diffuse gas in the outer reaches of the core do not grow much. This process of `competitive accretion' may be a major element in the origin of the IMF, as first pointed out by {\em Zinnecker} (1982).

Over the past decade, many numerical simulations of the formation of star clusters by fragmentation and competitive accretion have been performed (e.g. {\em Chapman et al.}, 1992; {\em Turner et al.}, 1995; {\em Whitworth et al.}, 1995; {\em Bonnell et al.}, 1997, 2001a, 2001b; {\em  Klessen et al.}, 1998; {\em Klessen and Burkert}, 2000, 2001; {\em  Bate et al.}, 2002a, 2002b, 2003; {\em Goodwin et al.} 2004b, 2004c). {\em Bonnell et al.} (1997, 2001a, 2001b) have shown through numerical simulations and analytical arguments that competitive accretion in large-$N$ proto-clusters can reproduce the general form of the IMF.

\subsection{\textbf{Unstable multiple systems}}

Multiplicity studies of main sequence and evolved stars have revealed that 15 to $25\%$ of all stars, when studied in sufficient detail, are triple or higher-order multiples (e.g. {\em Tokovinin and Smekhov}, 2002; {\em Tokovinin}, 2004). It follows that the formation of multiple stars is an important element of the star formation process.

The multiplicity fraction among PMS stars is poorly known, partly because of the difficulty in studying stars in the embedded phase, but it appears to be at least as high as -- and probably much higher than -- for more evolved stars (e.g. {\it Reipurth}, 2000; {\it Looney et al.}, 2000; {\em Koresko}, 2002; {\it Reipurth et al.}, 2002, 2004; {\it Haisch et al.}, 2004), see also the chapters by {\em Duch\^ene et al.}, {\em  Goodwin et al.} and {\em Burgasser et al.} in this volume. Most stars are formed in embedded clusters, and there is increasing evidence that the primary building blocks of clusters are small subclusters of $\la 20$ stars which quickly dissolve and merge to form the more extended cluster (e.g. {\em Teixeira et al.}, 2006).

Nonhierarchical multiple systems, in which the time-averaged distances between components are comparable, are inherently unstable (e.g {\em van Albada}, 1968). Within about a hundred crossing times a triple system is likely to have ejected one member, most likely the least massive component, since the ejection probability scales approximately as the inverse third power of the mass (e.g. {\em Anosova}, 1986; {\em Mikkola and Valtonen}, 1986). Although most nonhierarchical systems disintegrate in this way, the existence of numerous stable hierarchical triple systems shows that this is not always the case. Ejected members leave with a velocity that, to first order, is comparable to the velocity attained at pericenter in the close triple encounter, and depends on the geometry of the encounter, the energy and angular momentum of the system, and the masses of the components (e.g. {\em Standish}, 1972; {\em Monaghan}, 1976; {\em Sterzik and Durisen}, 1995, 1998; {\em Armitage and Clarke}, 1997).

The disintegration of a small multiple system is most likely to occur
during the deeply embedded Class 0 phase, while massive accretion from
a surrounding envelope is still taking place ({\em Reipurth 2000}). If
the ejection leads to an escape, then the accretion halts and the
final mass of the object is capped ({\em Klessen, Burkert, \& Bate
  1998}).

\subsection{\textbf{Dynamical ejection as a source of brown dwarfs}}

If a protostellar embryo is ejected from its natal core with a mass below the H-burning limit, then it becomes a brown dwarf ({\em Reipurth and Clarke}, 2001).  All that is needed for this to happen is for a prestellar core to spawn more than two protostellar embryos; for at least one of them to be less massive than $0.075\,{\rm M}_{_\odot}$ at the outset; and for one of them to stay less than $0.075\,{\rm M}_{_\odot}$ long enough to be ejected by dynamical interaction with the other embryos.

Simulations suggest that forming more than two protostellar embryos in a collapsing core is routine, as is the ejection of brown dwarfs and very low-mass stars (e.g.  {\em Bate et al.}, 2002a; {\em Delgado-Donate et al.}, 2003, 2004; {\em Goodwin et al.}, 2004a, 2004b).  However, these simulations may be misleading. (i) They use sink particles ({\em Bate, Bonnell and Price}, 1995), and thereby create protostellar embryos which at the outset are inevitably low-mass ($\sim 0.005\,{\rm M}_{_\odot}$), very prone to dynamical interaction (being effectively point masses, albeit with gravity softening), and unable to merge. Thus, although formation by ejection seems inevitable, it may be less efficient than these simulations suggest. As discussed by {\em Goodwin and Kroupa} (2005) and {\em Hubber and Whitworth} (2005), the observed binary statistics are incompatible with too many ejections, and the number of protostellar embryos undergoing dynamical interactions within a single prestellar core should be relatively small (${\cal N} \la 4$). (ii) They do not include magnetic fields. {\em Boss} (2002) argues that a magnetic field may promote fragmentation of a collapsing core, by inhibiting the formation of a central density peak; however, his code does not capture all the possible MHD effects, in particular the anisotropy of magnetic pressure and the torques exerted by a twisted field. In contrast {\em Hosking and Whitworth} (2004) have simulated the collapse of a rotating core with imperfect MHD, and find that fragmentation is inhibited. More work is needed.

\begin{figure}[h]
\epsscale{1.0}
\plotone{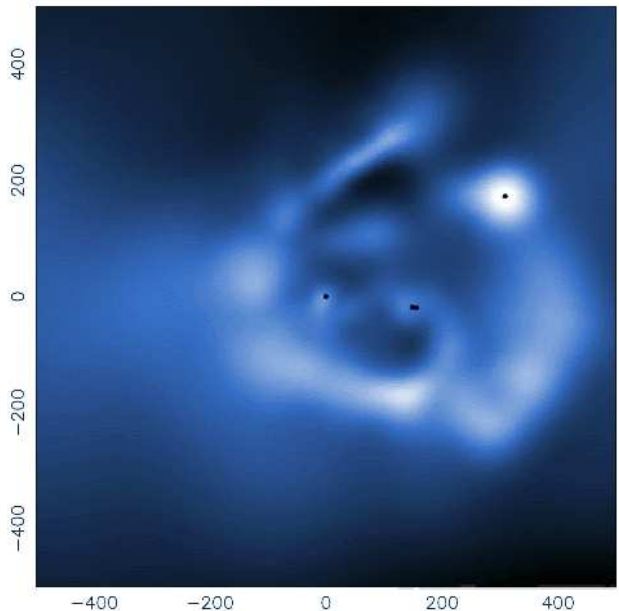}
\caption{\small Four stars formed in a low-mass core ($\sim 5\,{\rm M}_{_\odot}$) with a low initial level of turbulence ($E_{_{\rm TURB}} \sim 0.05\,|E_{_{\rm GRAV}}|$) ({\em Goodwin et al.} 2004b). Near the centre is a triple system containing a close binary (barely resolved pair of black dots). The object towards the upper righthand corner is a brown dwarf which has been ejected and is unbound from the triple. It has a significant disc ($M_{_{\rm DISC}} \sim 0.001\,{\rm M}_{_\odot}$). The frame is $10^3\,{\rm AU}$ across, and the time is $0.073\,{\rm Myr}$ since the start of collapse.}  
\end{figure}

In their original paper, {\em Reipurth and Clarke} (2001) conjectured that brown dwarfs formed by ejection could have a higher velocity dispersion than more massive H-burning stars, and that this might be detected, either in observations of the velocity dispersions of young star clusters, or as a diaspora of brown dwarfs around more evolved clusters. They also pointed out that violent ejections would result in smaller accretion discs, and therefore a shortened quasi-T Tauri phase. Numerical simulations of star cluster formation in highly turbulent massive cores (e.g. {\em Bate et al.}, 2003; {\em Bate and Bonnell}, 2005) find that the brown dwarfs do indeed have quite a high velocity dispersion ($\sim 2\;{\rm to}\;4\,{\rm km}\,{\rm s}^{-1}$), but they are difficult to distinguish from the low-mass H-burning stars, because both are frequently ejected from small dynamically-unstable groups. Simulations of low-density star-forming regions by {\em Goodwin et al.} (2005) find that the velocity dispersions are typically somewhat lower ($\sim 0.5\;{\rm to}\;1\,{\rm km}\,{\rm s}^{-1}$), and again the velocity dispersions of brown dwarfs and H-burning stars are hard to distinguish, because they are partially masked by the velocity dispersion between the different cores in which small subgroups ($\la 5$) of stars are born. For the same reasons, segregation of brown dwarfs from H-burning stars may be difficult to detect. Moreover, given their low ejection speeds, some brown dwarfs formed by ejection retain significant discs, and are therefore presumably well able to sustain accretion and outflows, as observed (see \S 2.3 and the chapter by Luhman et al.). Thus kinematical and spatial information on brown dwarfs, and the signatures of accretion and outflow, provide less powerful constraints on brown dwarf formation than were initially surmised by {\em Reipurth and Clarke} (2001).

\subsection{\textbf{Dynamical interactions in early stellar evolution}}

The binary statistics of brown dwarfs provide another potential constraint on formation mechanisms (see \S 2.2 and the chapters by {\em Burgasser et al.}, {\em Duch\^ene et al.}, {\em Goodwin et al.}, and {\em Luhman et al.}). Formation by ejection may be incompatible with the relatively high frequency of close BD-BD binaries, if this is confirmed (e.g. {\em Maxted and Jeffries}, 2005; {\em Joergens}, 2005a). As well as resulting in the ejection of some brown dwarfs, dynamical interactions must also influence the binary statistics (distributions of primary mass, mass ratio, semi-major axis and eccentricity) of the brown dwarfs and low-mass H-burning stars that do not get ejected (e.g. {\em Sterzik and Durisen}, 2003; {\em Kroupa and Bouvier}, 2003). {\em Hubber and Whitworth} (2005) have shown that if they take the observed distributions of core mass, radius and rotation rate, convert each core into a ring of 4 to 5 stars with masses drawn from a log-normal distribution having dispersion $\sigma_{_{\log_{10}[M]}} = 0.6$, and then follow the dissolution of the ring by pure $N$-body dynamics, they reproduce rather well the observed distribution of multiplicity and binary statistics in young clusters, as a function of primary mass.  {\em Umbreit et al.} (2005) have investigated the disintegration of nonhierarchical accreting triple systems, and find that they are able to produce the observed separation distribution of close binary brown dwarfs. Thus dynamical interactions may make important contributions, both to the formation of brown dwarfs, and to their binary statistics.

In addition to producing brown dwarfs and very low mass stars, and helping to shape the lower-mass end of the IMF and the statistics of binary and higher multiple systems, dynamical interactions may be the key to understanding a variety of other phenomena in early stellar evolution. {\em Reipurth} (2000) notes that the different sizes and separations of shocks in Herbig-Haro flows can be understood as a fossil record of the dynamical evolution of a newly formed binary. He also notes that dynamical interactions may on occasion lead to a departure from the standard evolutionary picture of a star passing smoothly from the Class 0 stage through the Class III stage. Instead stochastic dynamical interactions may lead to the sudden ejection of an object from the Class 0 or I stage, resulting in its abrupt appearance as a Class II or even a Class III object. The infrequent young binaries with infrared companions may be related to such events. Finally, the FU~Orionis eruptions may be related to the formation of a close binary, and could result from the viscous interactions of circumstellar material as the two components in a newly formed binary spiral together ({\em Bonnell and Bastien}, 1992; {\em Reipurth and Aspin}, 2004).

\section{\textbf{PHOTO-EROSION OF PRE-EXISTING CORES}} \label{SEC:EROS}%


A fifth -- and somewhat separate -- mechanism for forming brown dwarfs is to start with a pre-existing core of standard mass (i.e. $\la {\rm M}_{_\odot}$) and have it overrun by an HII region ({\em Hester et al.}, 1996). As a result, an ionisation front (IF) starts to eat into the core, `photo-eroding' it. The IF is preceded by a compression wave (CW), and when the CW reaches the centre, a protostar is created, which then grows by accretion. At the same time, an expansion wave (EW) is reflected and propagates outwards, setting up the inflow which feeds accretion onto the central protostar. The outward propagating EW soon meets the inward propagating IF, and shortly thereafter the IF finds itself ionising gas which is so tightly bound to the protostar that it cannot be unbound by the act of ionisation. All the material interior to the IF at this juncture ends up in the protostar. On the basis of a simple semi-analytic treatment, {\em Whitworth and Zinnecker} (2004) show that the final mass is given by
\begin{equation} \nonumber
\sim 0.01{\rm M}_{_\odot}
\left(\!\frac{a_{_{\rm I}}}{0.3\,{\rm km}\,{\rm s}^{-1}}\!\right)^6
\left(\!\frac{\dot{\cal N}_{_{\rm LyC}}}{10^{50}\,{\rm s}^{-1}}\!\right)^{-\frac{1}{3}}
\left(\!\frac{n_{_{\rm O}}}{10^3\,{\rm cm}^{-3}}\!\right)^{-\frac{1}{3}}\!,
\end{equation}
where $a_{_{\rm I}}$ is the sound speed in the neutral gas of the core, $\dot{\cal N}_{_{\rm LyC}}$ is the rate at which the exciting star(s) emit ionising photons, and $n_{_{\rm O}}$ is the density in the HII region.

This mechanism is rather robust, in the sense that it produces very low-mass stars for a wide range of initial conditions, and these conditions are likely to be realised in nature. Indeed, the evaporating gaseous globules (EGGs) identified in M16 by {\em Hester et al.} (1996) -- and subsequently in other HII regions -- would appear to be pre-existing cores being photo-eroded in the manner we have described (e.g. {\em McCaughrean \& Anderson}, 2002). However, the mechanism is also very inefficient, in the sense that it usually takes a rather massive pre-existing prestellar core to form a single brown dwarf or very low-mass H-burning star. Moreover, the mechanism can only work in the immediate vicinity of an OB star, so it cannot explain the formation of all brown dwarfs, and another mechanism is required to explain those seen in star formation regions like Taurus. Nonetheless, if the majority of stars are born in Trapezium like clusters, rather than Taurus-like regions, then photo-erosion should remain a contender for producing some brown dwarfs. Brown dwarfs formed by photo-erosion should include close BD-BD binaries. It is unclear whether they can retain significant accretion discs.

\section{\textbf{SIMULATIONS OF CLUSTER FORMATION}} \label{SEC:SIMS}%

With several different likely mechanisms for the production of brown dwarfs, the question arises: which, if any, is the dominant formation mechanism?  This is only likely to be answered through numerical simulations that are able to model the full star formation process, including all the relevant physical ingredients (gravity, hydrodynamics, magnetic fields, radiative transfer, and chemistry). There is a huge effort underway to perform such simulations, but there are formidable numerical challenges to overcome.

\subsection{Turbulent cloud collapse}

The most comprehensive simulations to date are those of {\em Bate, Bonnell and Bromm} (2002a, 2003), {\it Bate and Bonnell} (2005) and {\it Bate} (2005). These model the collapse and fragmentation of turbulent molecular clouds with mass $M \sim 50\,{\rm M}_{_\odot}$, initial diameter $0.18\;{\rm to}\;0.38\,{\rm pc}$ and initial temperature $T \sim 10\,{\rm K}$, to form small stellar clusters containing $\sim 50$ stars, including numerous brown dwarfs.  The clouds are seeded with a power spectrum of supersonic velocity structure that matches the scaling of velocity dispersion with length-scale observed in molecular clouds ({\it Larson}, 1981) and is allowed to decay during the simulations. The key difference between these simulations and earlier ones is that they are able to produce large numbers of objects from which statistical quantities can be derived (e.g. the form of the IMF), but simultaneously they also resolve down to the opacity limit for fragmentation (Section \ref{SEC:PHYS}) and so they are able to follow the formation of all the stars that the clouds produce.  They also resolve gaseous discs with radii down to $\sim 10\,{\rm AU}$ and binaries with separations greater than $\sim 1\,{\rm AU}$.  On the other hand, they do not include magnetic fields, radiative transfer, chemistry or feedback.  Therefore, for instance, they are unable to investigate the fraction of brown dwarfs that might form via photo-erosion.

\begin{figure*}[!ht]
\epsscale{2.0}
\plotone{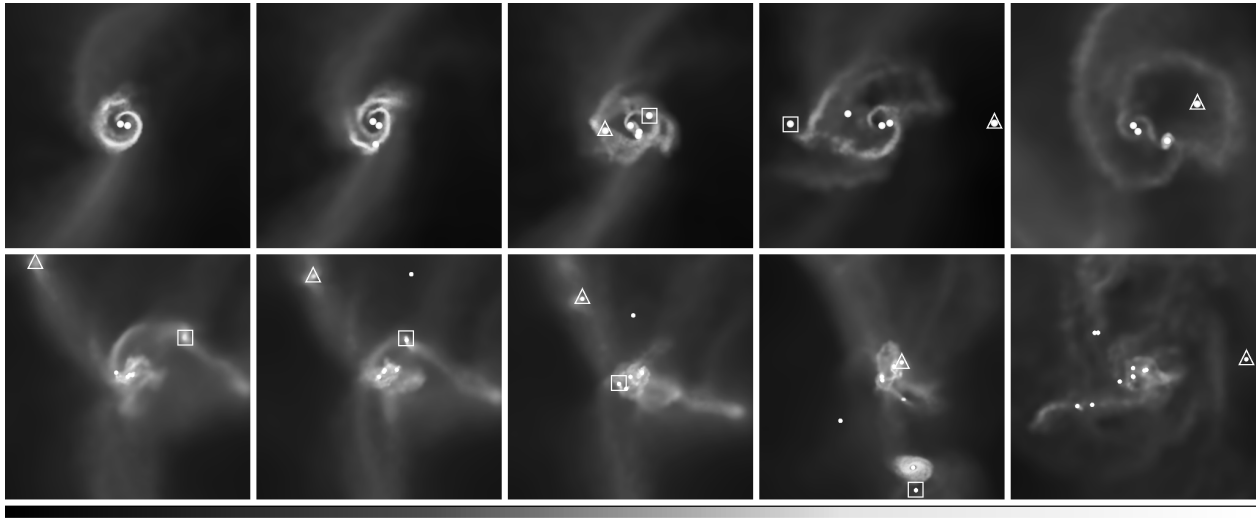}
\caption{\small Two sequences illustrating the brown dwarf formation mechanisms that occur in the simulations of Bate et al. (2002a, 2002b, 2003).  The upper sequence shows two brown dwarfs (square and triangle) forming in a circumbinary disc, while the lower sequence shows two brown dwarfs (square and triangle) forming in a filament. In both cases, these objects remain as brown dwarfs because they are dynamically ejected before they can accrete sufficient mass to ignite H-burning.  In the upper case, they are ejected from a multiple system formed by disc fragmentation.  In the lower case, the two brown dwarfs form in separate filaments, but then fall into, and are ejected from, the multiple system that exists at the intersection of the two filaments. Each panel is $600\,{\rm AU}$ across.}  
\end{figure*}

With these caveats in mind, the simulations generate clusters which are quite realistic. Starting with initial conditions typical of the molecular clouds in our Galaxy, the simulations produce roughly equal numbers of stars and brown dwarfs, with an IMF which is roughly compatible with that observed, at low masses; the high-mass end of the IMF ($\ga {\rm M}_{_\odot}$) is not usefully constrained by these simulations.  All stars, including brown dwarfs, originate from the fragmentation of dense filaments of molecular gas and from disc fragmentation, as illustrated in Fig. 3. Crucially, however, those that end up as brown dwarfs are those that avoid accreting large amounts of gas from the surrounding cloud.  They are able to avoid becoming stars because they are ejected dynamically from unstable multiple systems, thereby terminating their accretion. In general it is easier to form brown dwarfs by disc fragmentation, because the resulting protostellar embryos are then born in a dense multiple system, and they can therefore be ejected quickly, before acquiring too much mass. In contrast, protostellar embryos formed by filament fragmentation are born in relative isolation, and they first have to fall down the filament into a dense cluster before they are ejected; their final mass is therefore less likely to stay below the H-burning limit. Thus, these simulations support the ejection hypothesis for the origin of brown dwarfs, with disc fragmentation also playing an important role in generating dynamically unstable systems.

The brown dwarfs formed in these simulations do not frequently have companions.  With three individual calculations now published, the overall frequency of BD-BD and very low-mass binaries is $\sim 5\%$. Most of these systems are closer than $\sim 20$ AU. Sun-like stars with brown dwarf companions at  separations less than 10 AU are also very rare, at $\sim 2\%$.  Wider brown dwarf companions at hundreds or thousands of AU are much more common, although none of these systems have reached dynamic stability when the simulations are terminated.  Most of these results are consistent with current observations, with the exception of the BD-BD binary frequency, where the value from the simulations is about three times lower than the observed value ($\sim 15\%$). There are at least two possible reasons for this. First, most observed and calculated BD-BD binaries have separations $\la 20\,{\rm AU}$, but the simulations are unable to resolve discs $\la 10\,{\rm AU}$, so the discrepancy might be solved with better resolution.  If this is not the case, missing physics may be the problem (e.g. the effects of radiative transfer on disc fragmentation). Finally, only one brown dwarf that was ejected during any of the simulations had a resolved disc (radius $R \ga 10\,{\rm AU}$). This implies that brown dwarfs formed from highly turbulent cores should only have small discs. In contrast, the simulations of cores with low levels of turbulence performed by Goodwin et al. (2004a, 2004b) produce brown dwarfs with somewhat larger discs (see Fig. 2). This indicates that disc size may be a function of birth environment. Although many brown dwarfs are observed to have discs from their spectral energy distributions, the distribution of their sizes is not currently known.

\section{\textbf{CONCLUSIONS}} \label{SEC:CONC}%

Since the statistical properties of brown dwarfs appear to form a continuum with those of low-mass H-burning stars (\S \ref{SEC:STAR}), we have argued that brown dwarfs form like stars, that is to say, on a dynamical timescale and by gravitational instability, with an homogeneous initial elemental composition, the same as the interstellar medium from which they form. In this regard brown dwarfs are distinct from planets, which we define to be objects that form on a longer timescale, by the accumulation of a rocky core and -- if circumstances allow -- the subsequent acquisition of a gaseous envelope, leading to an initially fractionated elemental composition and a deficit of light elements.

We have evaluated the minimum mass for a brown dwarf (\S \ref{SEC:PHYS}) by considering the cooling required for a very low-mass prestellar core to condense out on a dynamical timescale. We have treated several different scenarios: hierarchical 3-D fragmentation; one-shot 2-D fragmentation of a shock-compressed layer; and fragmentation of a Toomre-unstable disc. All three cases yield values of $M_{_{\rm MIN}}$ in the range $0.001\;{\rm to}\;0.004\,{\rm M}_{_\odot}$ for contemporary star formation in the solar vicinity. This suggests that there may be some overlap between the range of masses occupied by stars and planets.  In hotter environments and at earlier epochs, $M_{_{\rm MIN}}$ was probably larger, and brown dwarfs were therefore less common. We also suggest that the thermodynamics of discs make it easier for proto brown dwarfs to condense out at large radii $\ga 100\,{\rm AU}$, and that these proto brown dwarfs may fragment to produce close BD-BD binaries, due to the dissociation of H$_{_2}$ and the opacity gap at $T \sim 2000\,{\rm K}$.

We have discussed five possible mechanisms for forming brown dwarfs. Turbulent fragmentation of molecular clouds may deliver prestellar cores of such low mass that their subsequent collapse (and possible fragmentation) can only yield brown dwarfs (\S \ref{SEC:TURB}). The collapse and fragmentation of more massive cores is likely to deliver protostellar embryos with a wide range of masses, many of which are too low to support hydrogen burning (\S \ref{SEC:FRAG}). Similarly, disc fragmentation is likely to deliver low-mass protostellar embryos (\S \ref{SEC:DISC}). These protostellar embryos may undergo competitive accretion (as a result of which some of them evolve to higher mass) and dynamical interactions (as a result of which some of them are ejected before they reach H-burning masses) (\S \ref{SEC:EJEC}). Lastly, cores which find themselves overrun by an HII region may be photo-eroded by the resulting ionisation front and end up spawning brown dwarfs (\S \ref{SEC:EROS}). None of these mechanisms is mutually exclusive, and in the most advanced simulations of cluster formation (\S \ref{SEC:SIMS}) collapse and fragmentation, disc fragmentation, competitive accretion and dynamical ejection all occur concurrently.

However, even these simulations do not capture all the deterministic physics. It requires a fully radiative, effectively inviscid, 3-D magneto-hydrodynamical simulation to evaluate properly the thermal effects which influence the minimum mass for star formation, the angular momentum transport processes which influence the binary statistics, and the $N$-body dynamics which influence the clustering properties of brown dwarfs. Work to develop and validate such codes is ongoing.

We believe that all the proposed mechanisms operate in nature, and that once they have been properly modelled, {\it the ultimate task will be to determine their relative contributions to the overall brown dwarf population}. These relative contributions may depend on environment, metallicity and epoch, and may therefore lead to local and/or cosmological variations in the ratio of brown dwarfs to H-burning stars, and in the binary and accretion statistics of brown dwarfs.

\bigskip

\centerline\textbf{ REFERENCES}
\bigskip
\parskip=0pt
{\small
\baselineskip=11pt

\refs Andr\'e Ph., Ward-Thompson D., and Barsony M. (1993) 
{\em Astrophys. J., 406}, 122-141.

\refs Andr\'e Ph., Ward-Thompson D., and Barsony M. (2000) 
{\em Protostars and Planets IV}, 59-96.

\refs Anosova J. P. (1986) 
{\em Astrophys. Spa. Sci., 124}, 217-241.

\refs Armitage P. J. and Clarke C. J. (1997) 
{\em Mon. Not. R. Astron. Soc., 285}, 540-546. 

\refs Arquilla R. and Goldsmith P. F. (1985) 
{\em Astrophys. J., 297}, 436-454.

\refs Bacmann A., Andr\'e Ph., Puget J.-L., Abergel A., Bontemps S., and Ward-Thompson D. (2000) 
{\em Astron. \& Astrophys., 361}, 555-580.


\refs Barranco J. A. and Goodman A. A. (1998) 
{\em Astrophys. J., 504}, 207-222.

\refs Basri G. and Mart\'in E. L. (1999) 
{\em Astron. J., 118}, 2460-2465.

\refs Bastien P. (1983) 
{\em Astron. \& Astrophys., 119}, 109-116.

\refs Bastien P., Arcoragi J.-P., Benz W., Bonnell I. A., and Martel H. (1991) 
{\em Astrophys. J., 378}, 255-265.

\refs Bate M. R. (2005) 
{\em Mon. Not. R. Astron. Soc., 363}, 363-378.

\refs Bate M. R. and Bonnell I. A. (2005) 
{\em Mon. Not. R. Astron. Soc., 356}, 1201-1221.

\refs Bate M. R., Bonnell I. A., and Price N. M. (1995) 
{\em Mon. Not. R. Astron. Soc., 277}, 362-376.

\refs Bate M. R., Bonnell I. A., and Bromm V. (2002a) 
{\em Mon. Not. R. Astron. Soc., 332}, L65-L68.

\refs Bate M. R., Bonnell I. A., and Bromm V. (2002b) 
{\em Mon. Not. R. Astron. Soc., 336}, 705-713.

\refs Bate M. R., Bonnell I. A., and Bromm V. (2003) 
{\em Mon. Not. R. Astron. Soc., 339}, 577-599.

\refs B\'ejar V. J. S., Mart\'in E. L., Zapatero Osorio M. R., Rebolo R., Barrado y Navacu\'es D., Bailer-Jones C. A. L., Mundt R., Baraffe I., Chabrier G., and Allard F. (2001)  
{\em Astrophys. J., 556}, 830-836.

\refs Boffin H. M. J., Watkins S. J., Bhattal A. S., Francis N., and Whitworth A. P. (1998)  
{\em Mon. Not. R. Astron. Soc., 300}, 1189-1204.

\refs Bonnell I. A. (1994) 
{\em Mon. Not. R. Astron. Soc., 269}, 837-848.

\refs Bonnell I. A. and Bastien P. (1992) 
{\em Astrophys. J., 401}, L31-L34.

\refs Bonnell I. A. and Bate M. R. (1994a) 
{\em Mon. Not. R. Astron. Soc., 269}, L45-L48.

\refs Bonnell I. A. and Bate M. R. (1994b) 
{\em Mon. Not. R. Astron. Soc., 271}, 999-1004.

\refs Bonnell I. A., Bate M. R., Clarke C. J., and Pringle J. E. (1997) 
{\em Mon. Not. R. Astron. Soc., 285}, 201-208.

\refs Bonnell I. A., Bate M. R., Clarke C. J., and Pringle J. E. (2001a)
{\em Mon. Not. R. Astron. Soc., 323}, 785-794.

\refs Bonnell I. A., Clarke C. J., Bate M. R., and Pringle J. E. (2001b)
{\em Mon. Not. R. Astron. Soc., 324}, 573-579.

\refs Boss A. P. (1986) 
{\em Astrophys. J. Suppl., 62}, 519-552.

\refs Boss A. P. (1996) 
{\em Astrophys. J., 468}, 231-240.

\refs Boss A. P. (2002) 
{\em Astrophys. J., 568}, 743-753.

\refs Boss A. P. (2004) 
{\em Astrophys. J., 610}, 456-463.

\refs Boss A. P., Fisher R. T., Klein R. I., and McKee C. F. (2000) 
{\em Astrophys. J., 528}, 325-335.

\refs Bouy H., Brandner W., Mart\'in E. L., Delfosse X., Allard F., and Basri G. (2003) 
{\em Astron. J., 126}, 1526-1554.

\refs Bouy H., Mart\'in E. L., Brandner W., and Bouvier J. (2005) 
{\em Astron. J., 129}, 511-517. 

\refs Boyd D. F. A. and Whitworth A. P. (2004) 
{\em Astron. Astrophys., 430}, 1059-1066.

\refs Brice\~no C., Luhman K. L., Hartmann L., Stauffer J. R., and Kirkpatrick J. D. (2002) 
{\em Astrophys. J., 580}, 317-335.

\refs Burgasser A. J., Kirkpatrick J. D., Reid I. N., Brown M. E., Miskey C. L., and Gizis J. E. (2003a) 
{\em Astrophys. J., 586}, 512-526.

\refs Burgasser A. J., Kirkpatrick J. D., Burrows A., Liebert J., Reid I. N., Gizis J. E., McGovern M. R., Prato L., and McLean I. S. (2003b) 
{\em Astrophys. J., 592}, 1186-1192.

\refs Burgasser A. J., Kirkpatrick J. D., and Lowrance P. J. (2005) 
{\em Astron. J., 129}, 2849-2855.

\refs Burkert A. and Bodenheimer P. (1996) 
{\em Mon. Not. R. Astron. Soc., 280}, 1190-1200.

\refs Burkert A. and Bodenheimer P. (2000) 
{\em Astrophys. J., 543}, 822-830.

\refs Burkert A., Bate M.R., and Bodenheimer P. (1997) 
{\em Mon. Not. R. Astron. Soc., 289}, 497-504.

\refs Cai K., Durisen R. H., Michael S., Boley A. C., Mej\'ia A. C., Pickett M. K., and D'Alessio P. (2005) 
astro-ph/0508354.

\refs Cha S.-H. and Whitworth A. P. (2003) 
{\em Mon. Not. R. Astron. Soc., 340}, 91-104.

\refs Chabrier G. (2003) 
{\em Publs. Astron. Soc. Pacific, 115}, 763-795.

\refs Chapman S. J., Pongracic H., Disney M. J., Nelson A. H., Turner J. A., and Whitworth A. P. (1992) 
{\em Nature, 359}, 207-210.

\refs Close L. M., Siegler N., Freed M., and Biller B. (2003) 
{\em Astrophys. J., 587}, 407-422.

\refs Curry C. L. (2002) 
{\em Astrophys. J., 576}, 849-859.

\refs Delgado-Donate E. J., Clarke C. J., and Bate M. R. (2003) 
{\em Mon. Not. R. Astron. Soc., 342}, 926-938.

\refs Delgado-Donate E. J., Clarke C. J., and Bate M. R. (2004) 
{\em Mon. Not. R. Astron. Soc., 347}, 759-770.

\refs Duquennoy A. and Mayor M. (1991) 
{\em Astron. Astrophys., 248}, 485-524.

\refs Elmegreen B. G. (2000) 
{\em Astrophys. J., 530}, 277-281.

\refs Fern\'andez M. and Comer\'on F. (2001) 
{\em Astron. Astrophys., 380}, 264-276.

\refs Gammie C. F. (2001) 
{\em Astrophys. J., 553}, 174-183.

\refs Gizis J. E., Kirkpatrick J. D., Burgasser A., Reid I. N., Monet D. G. Liebert, J., and Wilson J. C. (2001) 
{\em Astrophys. J., 551}, L163-L166.

\refs Gizis J. E., Reid I. N., Knapp G. R., Liebert J., Kirkpatrick J. D., Koerner D. W., and Burgasser A. J. (2003) 
{\em Astron. J., 125}, 3302-3310.

\refs Goodman A. A., Benson P. J., Fuller G. A., and Myers P. C. (1993) 
{\em Astrophys. J., 406}, 528-547.

\refs Goodwin S. P. and Kroupa P. (2005) 
{\em Astron. Astrophys., 439}, 565-569.

\refs Goodwin S. P., Ward-Thompson D., and Whitworth A. P. (2002) 
{\em Mon. Not. R. Astron. Soc,. 330}, 769-771.

\refs Goodwin S. P., Whitworth A. P., and Ward-Thompson D. (2004a) 
{\em Astron. Astrophys., 414}, 633-650.

\refs Goodwin S. P., Whitworth A. P., and Ward-Thompson D. (2004b) 
{\em Astron. Astrophys., 419}, 543-547.

\refs Goodwin S. P., Whitworth A. P., and Ward-Thompson D. (2004c) 
{\em Astron. Astrophys., 423}, 169-182.

\refs Goodwin S. P., Hubber D. A., Moraux E., and Whitworth A. P. (2005) 
{\it Astron. Nachrichten, 326}, 1040-1043.

\refs Haisch K.E., Greene T.P., Barsony M., and Stahler S.W. (2004) 
{\em Astron. J., 127}, 1747-1754.

\refs Hayashi C. and Nakano T. (1963) 
{\em Prog. Theor. Phys., 30}, 460-474.

\refs Hennebelle P., Whitworth A. P., Cha S.-H., and Goodwin S. P. (2004) 
{\em Mon. Not. R. Astron. Soc., 348}, 687-701.

\refs Hester J. J., Scowen P. A., Sankrit R., Lauer T. R., Ajhar E. A., et al. (1996) 
{\em Astron. J., 111}, 2349-2360.

\refs Hosking J. G. and Whitworth A. P. (2004) 
{\em Mon. Not. R. Astron. Soc., 347}, 1001-1010.

\refs Hoyle F. (1953) 
{\em Astrophys. J., 118}, 513-528.

\refs Hubber D. A. and Whitworth A. P. (2005) 
{\em Astron. Astrophys., 437}, 113-125.

\refs Inutsuka S.-I. and Miyama S. M. (1992) 
{\em Astrophys. J., 388}, 392-399.

\refs Jayawardhana R., Ardila D. R., Stelzer B., and Haisch K. E., Jnr. (2003)
{\em Astron. J., 125}, 1515-1521.

\refs Joergens V. (2006a) 
{\em Astron. Astrophys.} in press (astro-ph/0509134).

\refs Joergens V. (2006b) 
{\em Astron. Astrophys.} in press (astro-ph/0509462).

\refs Joergens V., Fern\'andez M., Carpenter J. M., and Neuh\"auser R. (2003) 
{\em Astrophys. J., 594}, 971-981.

\refs Johnson B. M. and Gammie C. F. (2003) 
{\em Astrophys. J., 597}, 131-141. 

\refs Johnstone D., Wilson C. D., Moriarty-Schieven G., \\
Giannakopoulou-Creighton J., and Gregersen, E. (2000) 
{\em Astrophys. J. Suppls., 131}, 505-518.

\refs Jones C. E., Basu S., and Dubinski J. (2001) 
{\em Astrophys. J., 551}, 387-393.

\refs Kenyon M. J., Jeffries R. D., Naylor T., Oliveira J. M., and Maxted P. F. L. (2005)
{\em Mon. Not. R. Astron. Soc., 356}, 89-106.

\refs Kirk H., Johnstone D., and Di Francesco J. (2006) 
{\em astro-ph/0602089}

\refs Kirk J. M., Ward-Thompson D., and Andr\'e Ph. (2005) 
{\em Mon. Not. R. Astron. Soc., 360}, 1506-1526.

\refs Klessen R. S. and Burkert A. (2000)
{\em Astrophys. J. Suppl., 128}, 287-319.

\refs Klessen R. S. and Burkert A. (2001)
{\em Astrophys. J., 549}, 386-401.

\refs Klessen R. S., Burkert A., and Bate M. R. (1998) 
{\em Astrophys. J. Letts., 501}, L205-L208.

\refs Koresko C.D. (2002) 
{\em Astron. J., 124,} 1082-1088.


\refs Kroupa, P. and Bouvier J. (2003) 
{\em Mon. Not. R. Astron. Soc., 346}, 369-380.

\refs Krumholz M. R., McKee C. F., and Klein R. I. (2005a) 
{\em Nature, 438}, 332-334.

\refs Krumholz M. R., McKee C. F., and Klein R. I. (2005b) 
{\em Astrophys. J., 618}, 757-768.

\refs Kumar S. S. (1963) 
{\em Astrophys. J., 137}, 1121-1125.

\refs Larson R. B. (1981) 
{\em Mon. Not. R. Astron. Soc., 194}, 809-826.

\refs Larson R. B. (1985) 
{\em Mon. Not. R. Astron. Soc., 214}, 379-398.

\refs Laughlin G. and Bodenheimer P. (1994) 
{\em Astrophys. J., 436}, 335-354.

\refs Lin C. C., Mestel L., and Shu F. H. (1965) 
{\em Astrophys. J., 142}, 1431-1446.

\refs Looney L.W., Mundy L.G., and Welch W.J. (2000) 
{\em  Astrophys. J., 529}, 477-498.

\refs Low C. and Lynden-Bell D. (1976) 
{\em Mon. Not. R. Astron. Soc., 176}, 367-390.

\refs Lucas P. W. and Roche P. F. (2000)  
{\em Mon. Not. R. Astron. Soc., 314}, 858-864.

\refs Lucas P. W., Roche P. F., and Tamura M. (2005) 
{\em Mon. Not. R. Astron. Soc., 361}, 211-232.

\refs Luhman K. L. (2004) 
{\em Astrophys. J., 617}, 1216-1232.

\refs Luhman K. L., Stauffer J. R., Muench A. A., Rieke G. H., Lada E. A., Bouvier J., and Lada C. J. (2003) 
{\em Astrophys. J., 593}, 1093-1115.

\refs Marcy G. W, and Butler R. P. (2000) 
{\em Publs. Astron. Soc. Pacific., 112}, 137-140.

\refs Masunaga H. and Inutsuka S. (1999) 
{\em Astrophys. J., 510}, 822-827.

\refs Maxted P. and Jeffries R. (2005) 
{\em Mon. Not. R. Astron. Soc., 362}, L45-L49.

\refs McCarthy C. and Zuckerman B. (2004) 
{\em Astron. J., 127}, 2871-2884.

\refs McCaughrean M. J. and Anderson M. (2002) 
{\em Astron. Astrophys., 389}, 513-518.

\refs McCaughrean M. J., Zinnecker H., Anderson M., Meeus G., and Lodieu, N.  (2002) 
{\em The Messenger, 109}, 28-36.

\refs Mikkola S. and Valtonen M.J. (1986) 
{\em Mon. Not. R. Astron. Soc., 223}, 269-278.

\refs Mohanty S., Jayawardhana R., Natta A., Fujiyoshi T., Tamura M., 
and Barrado y Navascu\'es D. (2004) 
{\em Astrophys. J., 609}, L33-L36.

\refs Monaghan J.J. (1976) 
{\em Mon. Not. R. Astron. Soc., 176}, 63-72.

\refs Moraux E., Bouvier J., Stauffer J. R., and Cuillandre J.-C. (2003)  
{\em Astron. Astrophys., 400}, 891-902.

\refs Motte F. and Andr\'e Ph. (2001) 
{\em Astron. Astrophys., 365}, 440-464.

\refs Motte F., Andr\'e Ph., and Neri R. (1998) 
{\em Astron. Astrophys., 336}, 150-172.

\refs Motte F., Andr\'e Ph., Ward-Thompson D., and Bontemps S. (2001) 
{\em Astron. Astrophys., 372}, L41-L44.

\refs Muench A. A., Alves J., Lada C. J., and Lada E. A. (2001) 
{\em Astrophys. J., 558}, L51-L54.

\refs Muzerolle J., Hillenbrand L., Calvet N., Brice\~no C., and Hartmann L. (2003) 
{\em Astrophys. J., 592}, 266-281.

\refs Muzerolle J., Luhman K. L., Brice\~no C., Hartman L., and Calvet N. (2005) 
{\em Astrophys. J., 625}, 906-912.

\refs Myers P. C. (2005) 
{\em Astrophys. J., 623}, 280-290.

\refs Myers P. C. and Benson P. J. (1983) 
{\em Astrophys. J., 266}, 309-320.

\refs Myers P. C. and Gammie C. F. (1999) 
{\em Astrophys. J., 522}, L141-L144.

\refs Myers P. C., Fuller G. A., Goodman A. A., and Benson P. J. (1991) 
{\em Astrophys. J., 376}, 561-572.

\refs Nakajima T., Oppenheimer B. R., Kulkarni S. R., Golimowski D. A., 
Matthew K., and Durrance S. T. (1995) 
{\em Nature, 378}, 463-465.

\refs Natta A. and Testi L. (2001) 
{\em Astron. \& Astrophys., 376}, L22-L25.

\refs Natta A., Testi L., Muzerolle J., Randich S., Comer\'on F., and 
Persi P. (2004) 
{\em Astron. Astrophys., 424}, 603-612.

\refs Nelson A. F., Benz W., Adams F. C., and Arnett D. (1998) 
{\em Astrophys. J., 502}, 342-371.

\refs Nordlund {\AA}. and Padoan P. (2003) 
{\em LNP Vol. 614: Turbulence and Magnetic Fields in Astrophysics}, 271-298.

\refs Onishi T., Mizuno A., Kawamura A., Tachihara K., and Fukui Y. (2002) 
{\em Astrophys. J., 575}, 950-973.

\refs Oppenheimer B. R., Kulkarni S. R., Matthews K., and Nakajima T. (1995) 
{\em Science, 270}, 1478-1479.

\refs Padoan P. and Nordlund {\AA}. (2002) 
{\em Astrophys. J., 576}, 870-879.

\refs Padoan P. and Nordlund {\AA}. (2004) 
{\em Astrophys. J., 617}, 559-564.

\refs Padoan P., Kritsuk A., Norman M. L., and Nordlund {\AA}. (2005) 
{\em Astrophys. J. Lett., 622}, L61-L64.

\refs Peng R., Langer W. D., Velusamy W. T., Kuiper T. B. H., and Levin S. (1998) 
{\em Astrophys. J., 497}, 842-849.

\refs Pinfield D. J., Dobbie P. D., Jameson R. F., Steele I. A., Jones H. R. A., and Katsiyannis A. C. (2003) 
{\em Mon. Not. R. Astron. Soc., 342}, 1241-1259.

\refs Preibisch T., McCaughrean M. J., Grosso N., Feigelson E. D., Flaccomio E., Getman K., Hillenbrand L. A., Meeus G., Micela G., Sciortino S., and Stelzer B. (2005) 
{\em Astrophys. J. Suppl., 160}, 582-593.

\refs Rafikov R. R. (2005) 
{\em Astrophys. J., 621}, L69-L72.

\refs Rebolo R., Zapatero Osorio M. R., and Mart\'in E. L. (1995) 
{\em Nature, 377}, 129-131.

\refs Rees M. J. (1976) 
{\em Mon. Not. R. Astron. Soc., 176}, 483-486.

\refs Reid M. A. and Wilson C. D. (2005) 
{\em Astrophys. J., 625}, 891-905.

\refs Reipurth B. (2000) 
{\em Astron. J., 120}, 3177-3191.

\refs Reipurth B. and Aspin C. (2004) 
{\em Astrophys. J., 608}, L65-L68.

\refs Reipurth B. and Clarke C. J. (2001) 
{\em Astron.\ J., 122}, 432-439.

\refs Reipurth B., Rodr\'\i guez L.F., Anglada G., and Bally J. (2002)
{\em Astron. J., 124}, 1045-1053.

\refs Reipurth B., Rodr\'\i guez L.F., Anglada G., and Bally J. (2004)
{\em Astron. J., 127}, 1736-1746.

\refs Rice W. K. M., Armitage P. J., Bate M. R., and Bonnell I. A. (2003) 
{\em Mon. Not. R. Astron. Soc., 339}, 1025-1030.

\refs Ryden B. S. (1996) 
{\em Astrophys. J., 471}, 822-831.

\refs Salpeter E. E. (1955) 
{\em Astrophys. J., 121}, 161-167.

\refs Sandell G. and Knee L. B. G. (2001) 
{\em Astrophys. J. Lett., 546}, L49-L52.

\refs Scholz A. and Eisl{\"o}ffel J. (2004) 
{\em Astron. Astrophys., 419}, 249-267.

\refs Slesnick C. L., Hillenbrand L. A., and Carpenter J. M. (2004) 
{\em Astrophys. J., 610}, 1045-1063. 

\refs Stamatellos D., Whitworth A. P., Boyd D. F. A., and Goodwin S. P. (2005) 
{\em Astron. Astrophys., 439}, 159-169.

\refs Standish E.M. (1972) 
{\em Astron. Astrophys., 21}, 185-191.

\refs Stassun K. G., Mathieu R. D., Vaz L. P. V., Valenti J. A., and Gomez Y. (2006) 
{\em Nature, }, in press.

\refs Sterzik M.F. and Durisen R.H. (1995) 
{\em Astron. Astrophys., 304}, L9-L12.

\refs Sterzik M.F. and Durisen R.H. (1998) 
{\em Astron. Astrophys.,  339}, 95-112.

\refs Sterzik M. and Durisen R. H. (2003) 
{\em Astron. Astrophys., 400}, 1031-1042.

\refs Tachihara K., Onishi T., Mizuno A., and Fukui Y. (2002) 
{\em Astron. Astrophys., 385}, 909-920.

\refs Teixeira P.S., Lada C.J., Young E.T. et al. (2006) 
{\em  Astrophys. J., 636}, L45-L48.

\refs Testi L. and Sargent A. I. (1998) 
 {\em Astrophys. J. Lett., 508}, L91-L94.

\refs Tokovinin A. A. (2004) 
in IAU Coll. 191 {\em The Environment and   Evolution of Double and Multiple Stars, Rev. Mex. Astron. Astrofis. Ser.Conf. 21,} 7-14.

\refs Tokovinin A. A. and Smekhov M.G. (2002) 
{\em Astron. Astrophys.,  382}, 118-123.

\refs Toomre A. (1964) 
{\em Astrophys. J., 139}, 1217-1238.

\refs Turner J. A., Chapman S. J., Bhattal A. S., Disney M. J., Pongracic H., and Whitworth A. P. (1995) 
{\em Mon. Not. R. Astron. Soc., 277}, 705-726.

\refs Umbreit S., Burkert A., Henning T., Mikkola S., and Spurzem R. (2005) 
{\em Astrophys. J., 623}, 940-951.

\refs van Albada T.S. (1968) 
{\em Bull. Astron. Inst. Netherlands, 19}, 479-499.

\refs Watkins S. J., Bhattal A. S., Boffin H. M. J., Francis N., and Whitworth A. P. (1998a) 
{\em Mon. Not. R. Astron. Soc., 300}, 1205-1213.

\refs Watkins S. J., Bhattal A. S., Boffin H. M. J., Francis N., and Whitworth A. P. (1998b) 
{\em Mon. Not. R. Astron. Soc., 300}, 1214-1224.

\refs Whelan E. T., Ray T. P., Bacciotti F., Natta A., Testi L., and Randich S. (2005) 
{\em Nature, 435}, 652-654.

\refs Whitehouse S. C. and Bate M. R. (2006) 
{\em Mon. Not. R. Astron. Soc.}

\refs Whitworth A. P. (2005)
{\em Astrophys. \& Space Sc. Library, 324}, 15-29.  

\refs Whitworth A. P. and Zinnecker H. (2004) 
{\em Astron. Astrophys., 427}, 299-306.

\refs Whitworth A. P., Bhattal A. S., Chapman S. J., Disney M. J., 
and Turner J. A. (1994a) 
{\em Mon. Not. R. Astron. Soc., 268}, 291-298.

\refs Whitworth A. P., Bhattal A. S., Chapman S. J., Disney M. J., 
and Turner J. A. (1994b) 
{\em Astron. Astrophys., 290}, 421-427.

\refs Whitworth A. P., Chapman S. J., Bhattal A. S., Disney M. J., Pongracic H., and Turner J. A. (1995) 
{\em Mon. Not. R. Astron. Soc., 277}, 727-746.

\refs Zapatero Osorio M. R., B\'ejar V. J. S., Marti\'in E. L., Rebolo R., Barrado y Navasqu\'es D., Mundt R., Eisl\"offel J., and Caballero J. A. (2002) 
{\em Astrophys. J., 578}, 536-542.

\refs Zinnecker H. (1982) 
{\em New York Academy Science Annals, 395}, 226-235.

\refs Zinnecker H. (1984) 
{\em Mon. Not. R. Astron. Soc., 210}, 43-56.

\end{document}